\documentclass[aps,pre,longbibliography,twocolumn,superscriptaddress,showpacs]{revtex4-1}

\usepackage{CJK}
\usepackage{graphicx}
\usepackage{amsmath}
\usepackage{dcolumn}
\usepackage{bm}
\usepackage[normalem]{ulem}
\usepackage{xcolor}
\usepackage{hyperref}

\usepackage{enumitem,kantlipsum}
\numberwithin{equation}{section}
 
\usepackage{epstopdf}
\newcommand{\rf}{\rho_{ref}}
\newcommand{\ms}{M_{\rm spinodal}}
\newcommand{\mb}{M_{\rm binodal}}

\newcommand{\rc}{\rho_{c}}
\newcommand{\tf}{\tilde{F}}
\newcommand{\tr}{\tilde{\rho}}
\newcommand{\tp}{\tilde{P}_0}

 \newcommand{\nn}{\nonumber}

\newcommand{\vb}{\langle\bv\rangle}

\newcommand{\bew}{\begin{widetext}}
\newcommand{\ew}{\end{widetext}}

\newcommand{\bq}{\mathbf{q}}
\newcommand{\bv}{\mathbf{v}}

\newcommand{\br}{\mathbf{r}}

\newcommand{\bqp}{\mathbf{q}_\perp}

\newcommand{\drho}{\delta v_x^{(1)}}

\newcommand{\bvp}{\mathbf{v}_\perp}

\newcommand{\hx}{\hat{x}}
\newcommand{\hp}{\hat{p}}

\newcommand{\bj}{\mathbf{j}}

\newcommand{\sep}{ \ \ \ , \ \ \ }

\newcommand{\beq}{\begin{equation}}
\newcommand{\eeq}{\end{equation}}
\newcommand{\beqn}{\begin{eqnarray}}
\newcommand{\eeqn}{\end{eqnarray}}
\newcommand{\pp}{\partial}

\newcommand{\partialder}[2]{\frac{\partial #1}{\partial #2}}

\begin{document}


\title{Spinodal decomposition and phase separation in polar active matter}
\author{Maxx Miller}
\email{maxxm@uoregon.edu}
\affiliation{Department of Physics and Institute for Fundamental
	 Science, University of Oregon, Eugene, OR $97403$}
\author{John Toner}
\email{jjt@uoregon.edu}
\affiliation{Department of Physics and Institute for Fundamental
	 Science, University of Oregon, Eugene, OR $97403$}

\begin{abstract}
We develop and study the hydrodynamic theory of flocking with autochemotaxis. This describes large collections of self-propelled entities all spontaneously moving in the same direction, each emitting a substance which attracts the others (e.g., ants). The theory combines features of the Keller-Segel model for autochemotaxis with the Toner-Tu theory of flocking. We find that sufficiently strong autochemotaxis leads to an instability of the uniformly moving state (the ``flock"), in which bands of different density form moving parallel to the mean flock velocity with different speeds. 
These bands, which are reminiscent of ant trails,  coarsen over time to reach a phase-separated state, in which one high density and one low density band fill the entire system.   The same instability, described by the same hydrodynamic theory, can occur in flocks  phase separating due to {\it any} microscopic mechanism (e.g., { sufficiently strong} attractive interactions).
 Although in many ways analogous to equilibrium phase separation via spinodal decomposition, the two steady state densities here are determined not by a  common tangent construction, as in equilibrium, but by an 
{\it un}common tangent construction very similar to that found for motility induced phase separation (MIPS) of disordered active particles. Our analytic theory agrees well with our numerical simulations of our equations of motion. 
\end{abstract} 
\maketitle

\section{Introduction and Motivation}

 Two of the principle mechanisms of {\it ordered} biological motion are ``taxis"
and ``flocking"\cite{Vicsek,TT1, TT2, TT3, TT4}. ``Taxis" is motion  in response to an external stimulus.  One  extremely large class of this type is movement directed preferentially along 
the gradient of some quantity. Examples include: ``thermotaxis'' (following temperature gradients)\cite{Thermotaxis} and ``phototaxis'' (light intensity)  
\cite{Phototaxis}. In this paper, we focus on the extremely common example of ``chemotaxis", in which motion is directed along the gradient of the concentration of a chemical substance. Almost all motile organisms perform chemotaxis \cite{chemotaxisExample}. 
Examples include: Escherichia coli\cite{Ecoli}, spongy moths\cite{Moth}, sperm\cite{Taxis1}, locusts \cite{Locust}, and ants\cite{Ants1, AntNoise1}.

 In ``autochemotaxis",  the chemical substance in 
question is secreted by the moving creatures themselves. Clearly, this can lead to aggregation 
of those creatures\cite{KS1, KS2, KS3, KSReview, KSCollective}.

Another ubiquitous, and familiar, form of motion is the phenomenon of flocking. Flocking is the collective coherent motion of a large number of organisms. Examples include: a school of fish, a swarm of bees, or a flock of birds. Such motion can occur any time creatures exhibit a tendency to follow their neighbors. Strikingly, even  purely short ranged interactions prove to be sufficient to generate long-ranged orientational order, even in an arbitrarily large flock, even in two dimensions\cite{TT1,TT2,TT3,TT4}, in contrast to equilibrium systems, in which the development of long-ranged orientational order in a rotation invariant system with purely short-ranged interactions is forbidden by the ``Mermin-Wagner-Hohenberg Theorem"\cite{MW}.

 Like chemotaxis, flocking can be observed on a large range of length scales from micrometers, as in, (e.g., slime molds \cite{Taxis3}, to kilometers (e.g., locust plagues) \cite{Locust}). Locust plagues can contain billions of locusts,  can occupy up to a thousand square kilometers, and are capable of causing billions of dollars of damage\cite{Locust}. 
 
 Hydrodynamic theories of autochemotaxis  without flocking\cite{KS1, KS2, KS3, KSReview, KSCollective} and flocking without autochemotaxis\cite{TT1, TT2, TT3,TT4, rean} have been successful in quantitatively explaining each of  their respective  phenomena in isolation. In this paper, we develop and study a hydrodynamic theory for systems in which {\it both} effects are present. We will show the presence of both effects can lead to formation of trails (e.g.,  ant trails\cite{Ants1}).
 
 Specifically, we consider active self-propelled particles (hereafter called ``boids")  that move over a frictional substrate, so that momentum is not conserved and time reversal symmetry is broken. The underlying dynamical rules are assumed to be rotation and translation invariant.

We investigate the stability in the presence of autochemotaxis of an ``active homogeneous polar ordered state" which spontaneously breaks the underlying rotation invariance by having all of the boids move, on average, in the same direction (i.e., ``flock"). The homogeneity of the state means that translation invariance is {\it not} broken.

The chemotaxic element of the hydrodynamic theory is the same as that of the Keller-Segel model\cite{KS1,KS2,KS3, KSReview}. Specifically, boids release a chemical signal. This chemical signal can be a substance that either repels neighboring boids (i.e., is a ``chemo-repellent") or attracts them (i.e., is a  ``chemo-attractant"). In this paper, we  study the chemo-attractant case, in which, in addition to following their neighbors, each boid also tends to follow the local chemo-attractant concentration gradient. The  chemo-attractant diffuses, and decays with a finite lifetime $\tau$.

The ``strength" of the autochemotaxis can therefore be increased by increasing any one of three parameters: the chemo attractant lifetime $\tau$, the rate of  production of the chemo attractant by the boids $\gamma$, or the coupling of their motion to the gradient of the concentration of the chemo-attractant.

\begin{figure}
    \centering
    \includegraphics[width=0.5\linewidth]{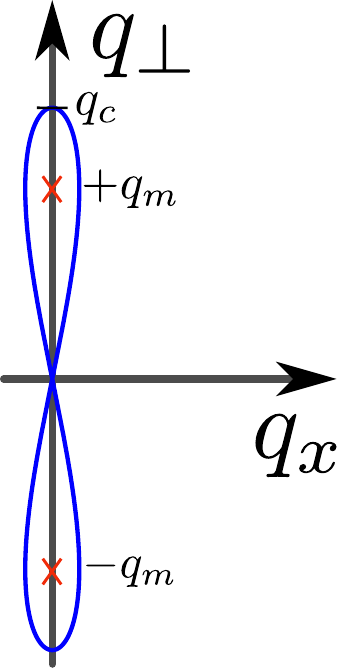}
    \caption{ The region of instability in $\bq$ space, for autochemotaxic systems very close to the instability threshold on the unstable side. Note that the instability only occurs at wavevectors very nearly perpendicular to the mean direction of flock motion, which is the horizontal ($x$)- axis in this figure. Specifically, the width along $q_x$ of the unstable region scales like  $\epsilon^2$, while its length perpendicular to $x$ scales like ${\epsilon}$, where $\epsilon$ is a measure of the distance from the instability threshold (defined more precisely below). As a result, the instability  is towards forming ``bands" in real space running {\it parallel} to the direction of mean flock motion, as illustrated in figures \ref{Band1} and \ref{Band2}.   }
    \label{q-unstable}
\end{figure}

The linear stability  analysis of the hydrodynamic theory that we perform in this paper shows  that, like disordered active systems undergoing only autochemotaxis \cite{KSCollective}, the spatially homogeneous state of ordered flocks (or, to use the technical term, ``polar ordered active fluids") becomes unstable when the autochemotaxis is sufficiently strong. However, in contrast to this instability in disordered systems, in ordered flocks the instability, near threshold, is extremely anisotropic. Specifically, plane wave modulations of the density with wavevector $\bq$ are only unstable in  a narrow  region of $\bq$ space, extended in the direction perpendicular to the direction of flock motion, as illustrated in figure \ref{q-unstable}. 

\begin{figure}
    \centering
  \includegraphics[width=\linewidth]{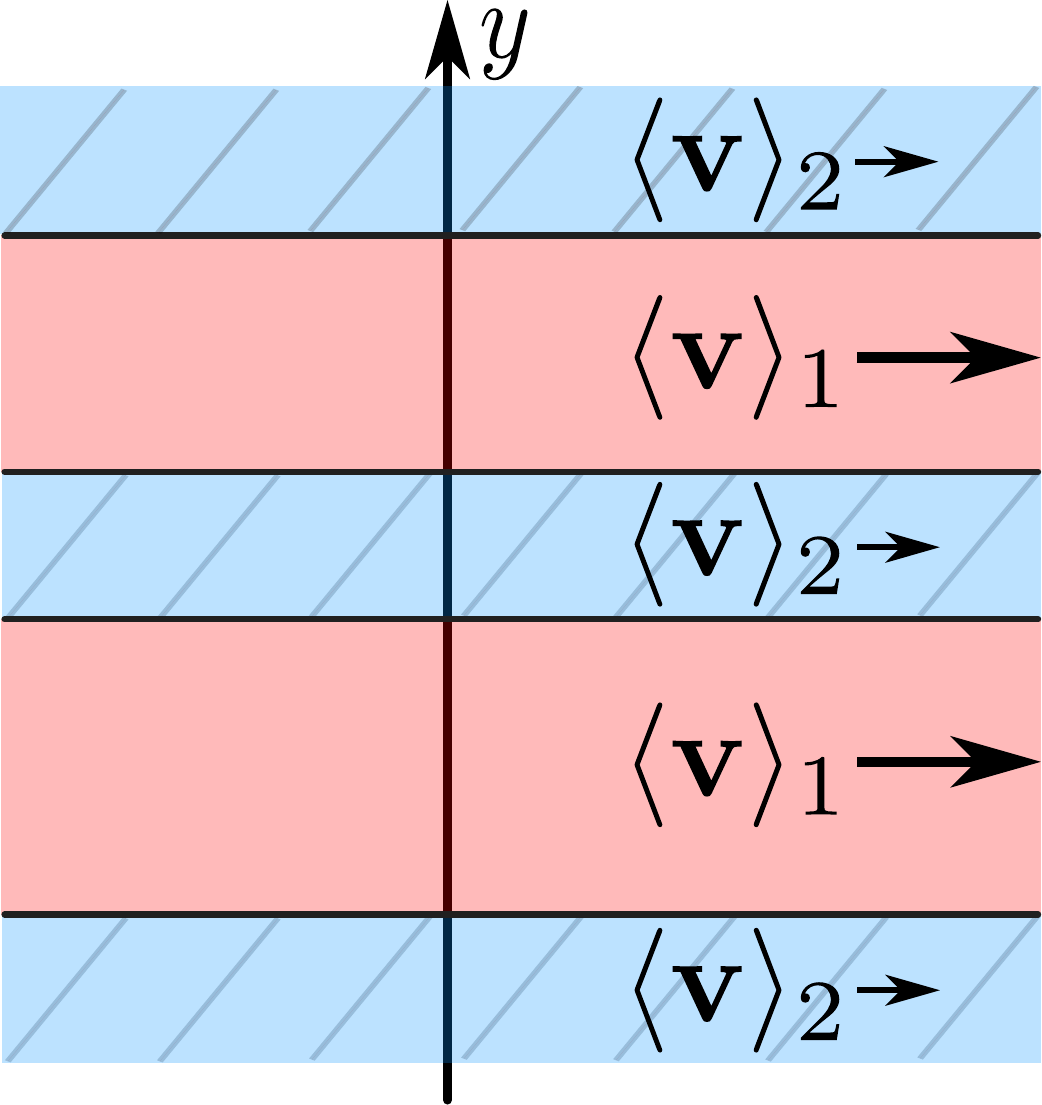}
    \caption{The ``band" structure of the instability at intermediate times. The density is only modulated along one of the directions (which we call $y$, and which is indicated in the figure) perpendicular to the direction $\hx$ of mean flock motion. A plot of the modulation of the density along the $y$-direction (e.g., along the path of the $y$-axis shown in the figure) is given on \ref{Band2}.}
    \label{Band1}
\end{figure}

\begin{figure}
    \centering
  \includegraphics[width=\linewidth]{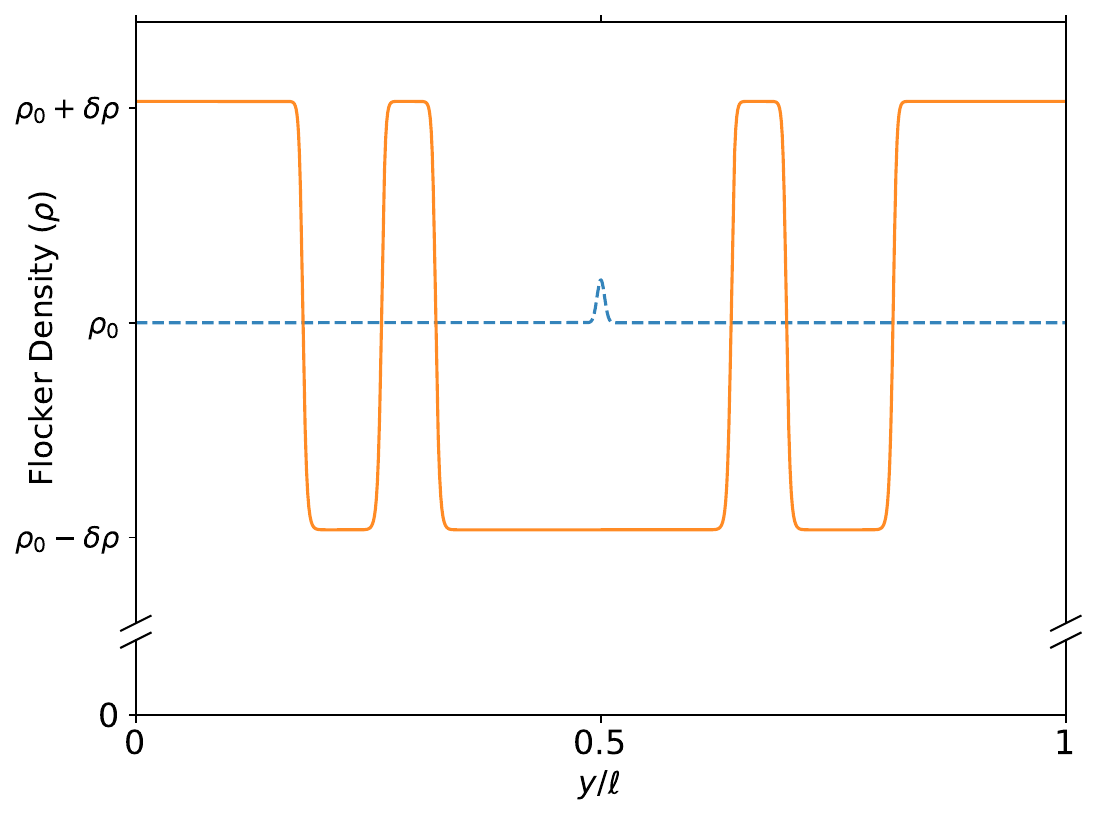}
    \caption{A plot of  the density for a typical numerical solution of the hydrodynamic equations for 1d configurations, as described in the text, at intermediate times, with weak non-linear terms.  The dashed blue curve is  the initial condition. The orange curve is the density at the last time step our simulation reached. The density is only modulated along one of the directions (which we call $y$, and which is indicated in the figure) perpendicular to the direction $\hx$ of mean flock motion. Here $\ell$ is the linear spatial extent of  our system in the $y$-direction (which is periodic). The parameters used for this solution were $\Gamma,\Lambda=0.25$, and the solution was iterated for 3000 units of time. The choice of $\rho_0$ was arbitrary.
    }
    \label{Band2}
\end{figure}

As a result, the instability is towards forming ``bands" in real space running {\it parallel} to the direction of mean flock motion. This is illustrated in figures \ref{Band1} and \ref{Band2}.

{  The instability we find here is completely different from the ``banding instability" \cite{BLee, banding1, Banding2, rean}. The banding instability occurs in a completely different region of parameter space\cite{rean}, is driven by completely different physics\cite{BLee, banding1, Banding2, rean}, and is characterized by the formation of bands {\it perpendicular} to the direction of mean flock motion.}

To investigate what happens once the density and velocity perturbations induced by this instability became too large to be treated by our linear stability analysis, 
we found  both analytic steady-state and numerical time-dependent solutions to our equations of motion. We did so for  ``one-dimensional" (1d)
solutions: that is, solutions in which the fields only depend on time and a single Cartesian coordinate perpendicular to the direction of mean flock motion. This one-dimensional restriction is justified, at least in the early stages, by the aforementioned fact that the instability only occurs for modes with their wavevectors nearly perpendicular to the direction of mean flock motion. 

\begin{figure}
    \centering
    \includegraphics[width=\linewidth]{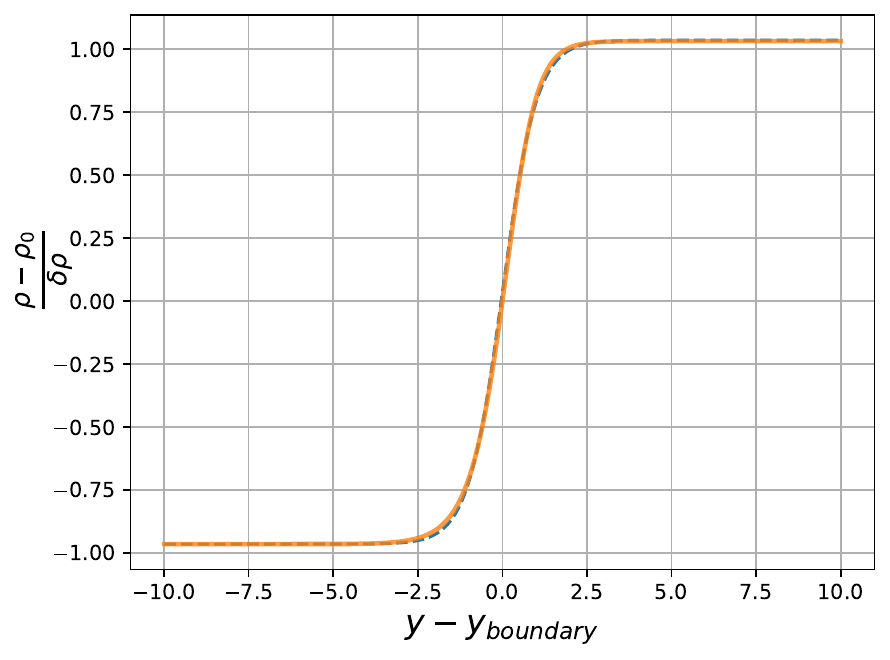}
    \caption{The density profile  of the final phase separated steady state. The long-time limit of our numerical solution, and our  fit curve are both plotted. The solid orange curve is the numerical solution. The dashed blue curve is the fit curve. The fit curve is $\tanh{(y/\sqrt{1.0263})} + 0.0355$. Although the form of the fit curve is not derived analytically, its large and small $y$ limits are taken from our analytic theory. As can be seen,  the agreement between the  analytic theory and the numerical solution  is good. The parameters used are $\frac{1}{\rho_0}\sqrt{\frac{m}{u}},\frac{\lambda}{D_{_{L\perp}}\rho_0}\sqrt{\frac{m}{u}}=0.25$. }
    \label{Band3}
\end{figure}

Our numerical solutions show that the
these bands fairly quickly evolve into a set of well-separated parallel ``plateaus'' of almost constant density, as illustrated in figure \ref{Band2}. The evolution then continues via the merger of high density bands with other high density bands and low density bands with  other low density ones. This suggests that bands of like density are attracted to each other;
however, it appears that this interaction falls off quickly with distance, because the merging process gets very slow.  Indeed, the mergers become  so slow  that  our 1d simulations simply cannot run long enough (at least in the time we're willing to wait!) to reach what we believe will be the ultimate steady state of the system. That state is  predicted by our analytic solution to be  phase separation into one high density band and one low density band, moving parallel to each other (and in the direction of the original flock motion) at slightly different speeds. This final geometry is illustrated in figure \ref{Band3},  in which we compare our analytic solution of the steady state equations to the numerical, time-dependent solution at a sufficiently late time.

Our argument for this final state is based on our analytic steady state solution of our equations of motion, which is illustrated in figure \ref{Band3}. This solution, which is asymptotically valid sufficiently close to the analog in our problem of the critical point in equilibrium phase separation, has all of the features of the density profile of  an equilibrium system in vapor-liquid coexistence: on the left we have a low density ``vapor" phase. As we move to the right, the density increases smoothly through an interface with a well-defined width, until it turns over and plateaus at a higher ``liquid" density. The approach to the liquid density as one moves to the right of the interface, and the approach to the vapor density as one moves to the left, is exponential. 

It is this exponential tail - or, more precisely, the overlap of the exponential tail of one high density region with that of one on the other side of an intervening low density region- that leads to the attractive interaction between bands. It is the fact that this interaction falls off exponentially that makes the merger process so slow. Indeed, we expect that the time to reach the fully phase separated state will, in this 1d picture, grow exponentially with the system size. We believe this is the reason the system cannot reach the final phase separated state in our  numerical solutions.

However, we also believe that this exponential growth of equilibration time with length scale is an artifact of the 1d nature of the analysis just presented. While we believe this approach is accurate for early times - i.e., times up to and including the formation of the plateaus- we also believe that once the plateaus are formed, their merger will be dominated by processes that are missed in a 1d picture.

\begin{figure}
    \centering
   \includegraphics[width=\linewidth]{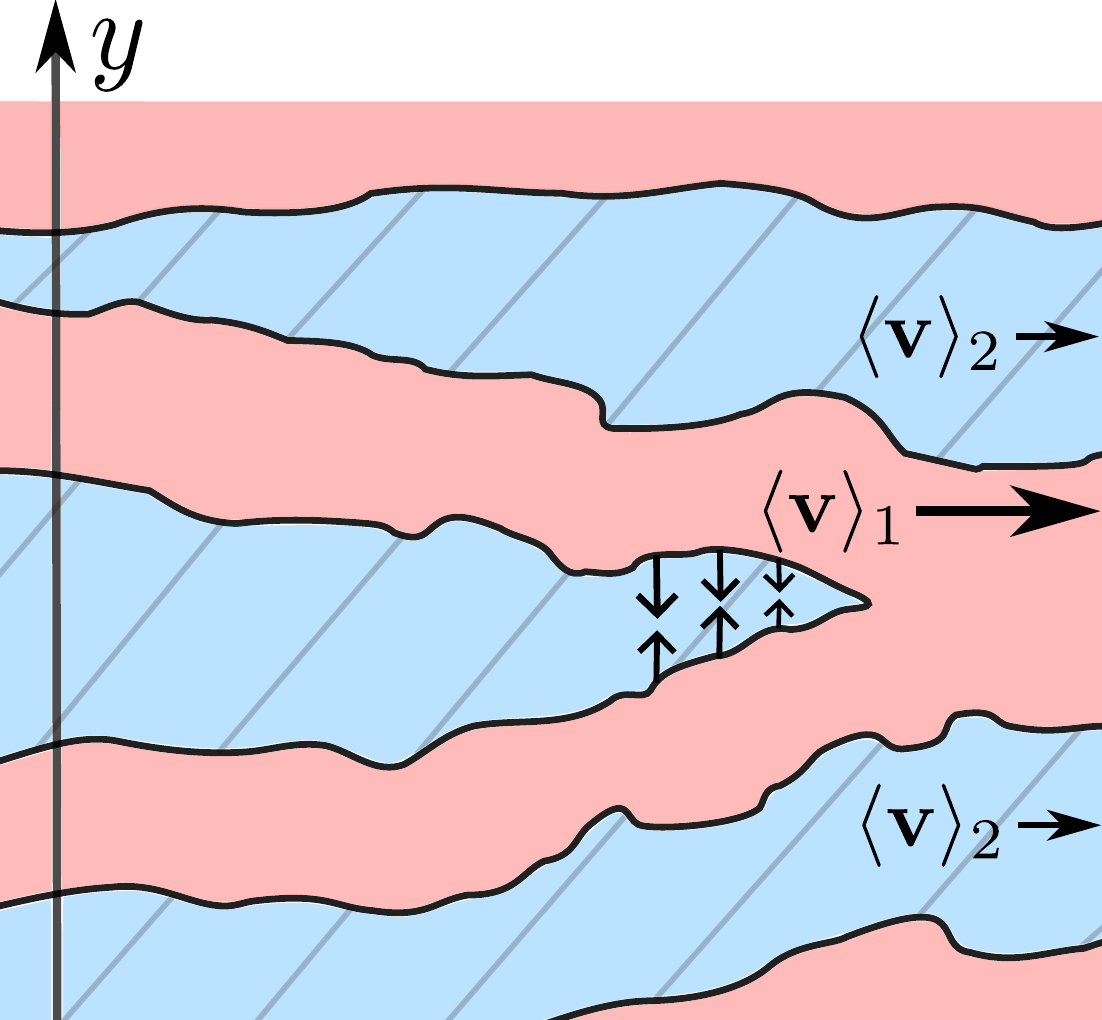}
    \caption{Our conjectured picture of the  evolution of the density bands  at later times. The boundaries fluctuate due to noise in ways that are not captured in the noiseless one dimensional approach. The central blue band is depicted in our conjectured transient ``zipper'' state.}
    \label{noiseBands}
\end{figure}

Specifically, we expect that what will really happen, once the plateaus form, is that the bands will begin to undulate due to noise in the system, as illustrated in figure \ref{noiseBands}. These undulations will grow with time until bands begin bumping into  their neighbors, at which point the bands can start to merge. The merged region will then rapidly ``zipper" along in both directions, merging the two bands.  This process will then repeat with the larger bands formed as a result of such mergers, until full phase separation is achieved.

The time scale needed for undulations of the bands to grow large enough to reach neighboring bands probably grows only {\it algebraically} with the separation between bands - much as the interface fluctuations in the KPZ equation do\cite{KPZ1} - rather than exponentially. We therefore expect that the time required for our autochemotaxic instability to completely phase separate a large system will grow only algebraically with system size.

However, to make this idea more precise, a theory of the dynamics of fluctuations of an interface between two flocks of different densities and speeds is needed. No such theory currently exists; its development remains  a topic for future research.

Those familiar with the dynamics of equilibrium phase separation will see the similarities with our system, although ours differs from the equilibrium case because of its strong anisotropy, and  non-equilibrium effects.

One manifestation of these non-equilibrium effects is the determination of  the values of the two co-existing steady-state densities. In equilibrium, these can be determined by the well-known ``common tangent construction"\cite{chaikin}. Our aforementioned analytic steady-state solution shows that in flocks that phase separate, the two coexisting densities are determined by an {\it  uncommon} tangent construction. This is very similar to the behavior found in mobility induced phase separation (MIPS) in {\it disordered} active systems \cite{mipsreview}. 

More specifically, we find that as one approaches the dynamical critical point, the uncommon tangent construction approaches a common tangent construction - but {\it only} in this limit. In general, the uncommon tangent construction applies.

\begin{figure}
        \centering
        \includegraphics[width=\linewidth]{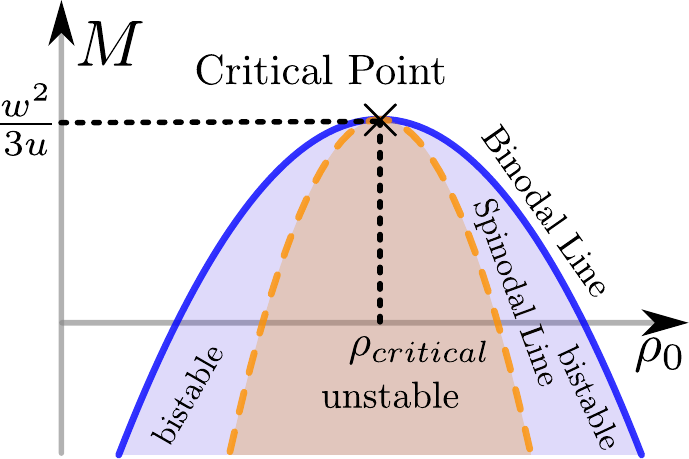}
        \caption{Phase diagram of flock phase-separation. Solid blue and dashed orange lines are the binodal and spinodal lines respectively. The orange filled region, labeled ``unstable", corresponds to the coexistence of the `liquid' (high density) and `gas' (low density) phases. See figure \ref{spinodal} and section V for more details.}
        \label{spinodalIntro}
\end{figure}

Our analytic steady-state solution  implies the phase diagram shown in figure \ref{spinodalIntro} and \ref{spinodal}, in which the vertical axis is a model parameter defined later which decreases with increasing autochemotaxis , and the horizontal axis is the mean density of the system. This analytic solution is asymptotically valid sufficiently close to 
the critical point in  this figure. In the region above the curve labelled ``Binodal line", the uniform state is the only stable steady state of the flock. Below the curve labelled ``Spinodal line", only the phase separated state is stable. Between these two curves, {\it either} state can be stable.

When the instability is weak, and when the system is close to  the dynamical analog of  an equilibrium liquid-gas critical point\cite{chaikin}, we can analytically determine the two densities in terms of parameters of the model. This analytic theory agrees extremely well with numerical solutions of our equations of motion, even when the instability is not weak and the system is fairly far from a critical point.

This said, our treatment of this phase separation is entirely ``mean-field": that is, it ignores fluctuations in the local density and velocity. Fluctuations in the density are well-known\cite{chaikin, Ma} to radically change the scaling behavior of the density near the critical point in equilibrium systems. Furthermore, even  stable flocks experience enormous (indeed, divergent) fluctuation corrections to their dynamics\cite{TT1,TT2,TT3,TT4}.  Among other effects, these make the diffusion ``constants" no longer constant; they become functions of the wave vector $\bq$ under consideration, diverging as $\bq\to{\bf 0}$. Since {\it both} types of effects - i.e., critical and non-critical fluctuations- are present in our system, we expect the scaling exponents for, e.g., the density difference between the two phase separated regions, to be quantitatively different from those we find here. Therefore, the   effect of fluctuations on the non-equilibrium phase separation we study here is a promising  open topic for future research.

Although we have focused in the above discussion on autochemotaxis, { the ultimate, longest length and time scale limit of} our model, and the  results we obtain for it herein,  also apply to {\it any}  dry polar { ordered} active matter { system in which {\it any} microscopic mechanism drives the inverse compressibility negative}. 
There are many other mechanisms one could imagine that could { do this: direct attractive interactions between the active agents, for one, would be sufficient}. { {\it Any} system in which {\it any} microscopic mechanism leads to a negative inverse compressibility will be described, at the longest length and time scales, by a Toner-Tu model with a negative inverse compressibility, which is the basis for all of our predictions here. Therefore, }, the phenomenology we predict here - in particular, that the phase separation is into bands of alternating  higher and lower density form running {\it parallel} to the direction of mean flock motion, as illustrated in figure \ref{Band1}-
will hold for {\it any} mechanism of phase separation (e.g., { sufficiently strong} attractive interactions between the boids).

The remainder of this paper is organized as follows: in section (\ref{review}), we review the Keller-Segel model of autochemotaxis\cite{KS1, KS2, KS3, KSReview, KSCollective}and the hydrodynamic (Toner-Tu) theory of flocking\cite{TT1, TT2, TT3, TT4, rean}. In section III, we develop the equations of motion for autochemotaxic flocks. In section IV, we linearize the equations of motion, examine their mode structure, and identify the instability. In section V, we analytically and numerically determine the final state to which the instability involves, using a one-dimensional analysis for systems very near a ``critical point", analogous to the critical point of phase separating equilibrium systems, that we identify. This proves to be described by a common tangent construction very similar to that for equilibrium systems. In section VI, we demonstrate that further from the critical point, this phase separation is described by an ``uncommon tangent construction" quite similar to that found in {\it disordered} active systems (e.g., motility-induced phase separation (MIPS)\cite{mipsreview}. In section VII, we  summarize our results, and discuss possible directions for future research.

\section{Review of prior theories of (auto)chemotaxis and flocking}{\label{review}}

\subsection{(Auto)chemotaxis}

In the 1970s, Keller and Segel developed their model of chemotaxis\cite{KS1, KS2, KS3, KSReview, KSCollective}.  The general model includes {\it auto}chemotaxis, which describes a collection of creatures (which we'll call ``boids") emitting  ``chemo-attractant"; i.e., a substance to which they themselves are attracted. The chemo-attractant then diffuses, and decays with a finite lifetime. The boids diffuse as well, but with a bias in the direction of the gradient of the chemo-attractant concentration. In addition, Keller and Segel in their most general formulation allow for birth and death of the boids,  making their system what we would now describe as a ``Malthusian flock"\cite{TT5, Malthus}. 

All of these processes (diffusion, chemo-attractant generation and decay) will in general depend on the local number density $\rho(\br, t)$ of  boids  at position $\br$  and time $t$,  and the local concentration $\eta(\br, t)$ of the chemo-attractant  at $(\br,t)$.
This reasoning lead Keller and Segel to the following  hydrodynamic equations in their most general form:
\beqn
    \pp_t\rho&=& \nabla \cdot\bigg( k_1(\rho, \eta) \nabla \rho - k_2(\rho,\eta)\rho\nabla \eta\bigg) + k_3(\rho, \eta)\,,\nn\\\label{KSrho}
    \\ 
    \pp_t\eta&=& D_\eta\nabla^2\eta + k_4(\rho, \eta) - k_5(\rho, \eta)\eta \,. \label{KSC}
\eeqn
In this expression, $k_1$ is the motility of the boids, $k_2$  the sensitivity to the chemical signal, $k_3$  the  difference between the boid birth and death rate, and  $k_4$ and $k_5$ give respectively the production and decay of the chemical signal. \cite{HillenPainterCompsci}

When $k_2$ is positive, the chemical signal acts as a chemo-attractant. When $k_4$ is positive and depends on $\rho$ (e.g., $k_4\propto\rho$),  the system is autochemotaxic. We will focus in this paper on flocks without birth and death, which amounts to taking $k_3=0$.  Otherwise, we will take the very general form (\ref{KSC}).

\subsection{Flocking}

Self-propelled entities in large numbers can display coherent motion over a wide range of length scales. Inspired by the work of Viscek and the historical progression of hydrodynamics, a hydrodynamic theory of flocking was developed by Toner and Tu. \cite{TT1, TT2, TT3, TT4, rean}.

The theory describes a flock composed of active, self-propelled particles moving over a frictional substrate, so that momentum is not conserved and time reversal symmetry is broken. In the jargon of the active matter field, this is called a ``dry" flock. The underlying dynamical rules are assumed to be rotation and translation invariant. These symmetries, together with the requirement that the total number of birds is conserved (that is, we ignore birth and death ``on the wing") strongly constrain the form of the theory.  In particular, they determine which variables we need to keep in a long-wavelength, long-time theory (which is what we mean by a ``hydrodynamic" theory).

In all hydrodynamic theories, it is sufficient to keep only the ``slow" variables; that is, the variables which evolve slowly at long length scales. More precisely, this means the variables whose time scales of evolution diverge as the length scale under consideration does. 

There are only two generic reasons that a variable will be slow in this sense: it must either be associated with a conserved quantity like  boid number, or be a ``Goldstone mode" arising from a broken continuous symmetry. For a dry flock as defined above,  the only conserved quantity is boid number; the associated hydrodynamic field is the coarse grained number density of boids $\rho(\br,t)$. Since an ordered flock spontaneously breaks the continuous rotational symmetry of the underlying dynamics, the boid velocity field $\bv(\br,t)$ (or, more precisely, its components perpendicular to the mean velocity of the ordered state) is a Goldstone mode, and hence, also slow.  

The equations of motion for these fields keep all terms that are ``relevant", in the renormalization group sense of affecting the scaling of the system at long distances and times, and that are consistent with the symmetry  (rotation invariance) and conservation law (boid number) of the system. The requirement of ``relevance" is largely met by performing a gradient expansion. As shown by  \cite{TT1,TT2,TT3,TT4}, this forces the equations of motion to be of the form:
\bew
\begin{align}
    \nonumber &\pp_t\bv+ \lambda_1(\bv\cdot\nabla)\bv + \lambda_2(\nabla\cdot\bv)\bv + \lambda_3\nabla(|\bv|^2) = \\
    &\quad \quad U(|\bv|,\rho)\bv - \nabla P_1(|\bv|,\rho) - \bv(\bv\cdot\nabla P_2(|\bv|,\rho)) + D_B\nabla(\nabla\cdot\bv) + D_T\nabla^2\bv + D_2(\bv\cdot\nabla)^2 \bv \label{TTv}\\
    &\pp_t\rho + \nabla\cdot(\bv\rho) = 0 \label{TTrho}
\end{align}
\ew

The significance of each of the terms in these equations is as follows:

The terms involving the parameters $\lambda_i$ are analogs of the convective derivative of the coarse grained velocity field in the Navier-Stokes equations. If our system respected Galilean invariance, we would have $\lambda_1 = 1$ and $ \lambda_{2,3} = 0$. However, because our flock is on a frictional substrate, which provides a special reference frame, we have neither Galilean invariance nor momentum conservation.

\begin{figure}
    \centering
    \includegraphics[width=\linewidth]{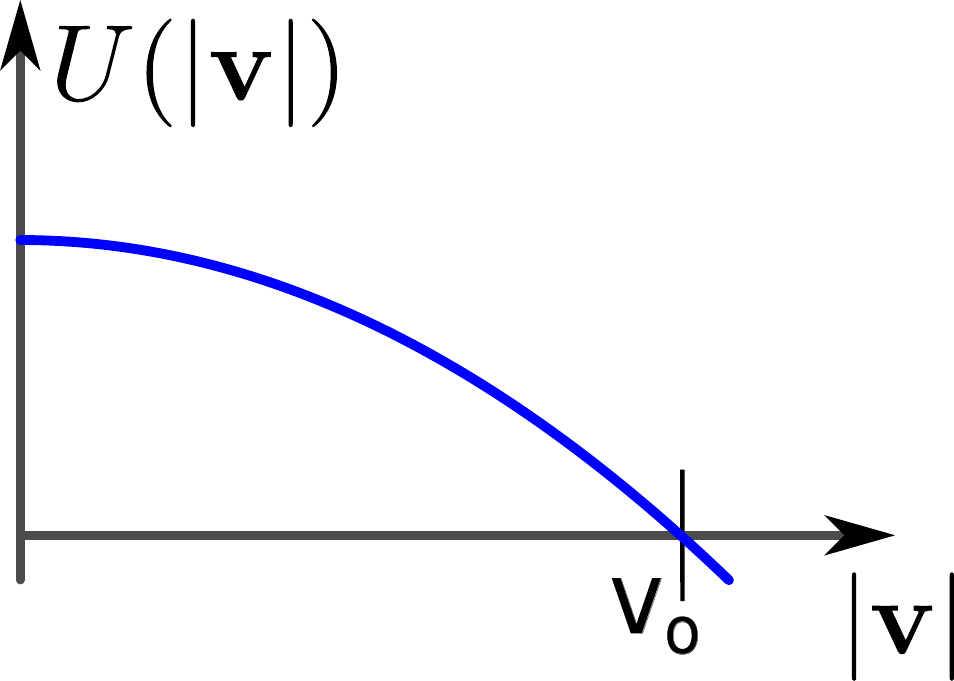}
    \caption{ A plot of $U(|\bv|, \rho)$ for fixed $\rho$. The effect of this term is to maintain a non-zero speed, $v_0$, of the flock. Boids will accelerate if their speed is below $v_0$ and decelerate if their speed is larger than $v_0$.}
    \label{disspationgraph}
\end{figure}

The $U(|\bv|,\rho)$ term, which is similar in form to  a dissipative term, clearly therefore also breaks both Galilean invariance  and momentum conservation. However, because our system is active, $U(|\bv|,\rho)$ need not be negative for all $|\bv|$. Indeed, if we are to model a system in which the steady state is a {\it moving} flock, we must take it to have the form illustrated in figure  \ref{disspationgraph}. In earlier literature\cite{TT1,TT2,TT3,TT4,TT5}, the special choice $U(|\bv|,\rho)=\alpha - u|\bv|^2$ is often made. This is by no means necessary, however.  Therefore, here we will make no assumptions about the precise form of $U(|\bv|,\rho)$, other than that it is analytic in $|\bv|$ and $\rho$, and that it has the qualitative shape illustrated in (\ref{disspationgraph}).
\vspace{.2in}

The $P_1$ term is perfectly analogous to  the isotropic pressure in the Navier-Stokes equations. The $P_2$ term is an ``anisotropic pressure", and is allowed because our system breaks rotation invariance locally, which means that there is no reason that the response to density gradients {\it along} the local velocity $\bv$ should be the same as that to gradients perpendicular to $\bv$. Note that this term also breaks Galilean invariance.

The  velocity diffusion constants $D_B$ and  $D_T$ are precise analogs of the bulk and shear viscosities, respectively, in the Navier-Stokes equations. The diffusion constant $D_2$ is an anisotropic viscosity which has no analog in the Navier-Stokes equation, because it violates Galilean invariance. Since we lack Galilean invariance here, it is allowed in our problem.
All of these viscosities have the effect of suppressing fluctuations of the velocity away from spatial uniformity.

The quantities $P_{1,2}(|\bv|,\rho)$, $U(|\bv|,\rho)$, $D_{B,T,2}$, and $\lambda_{1,2,3}$ are in general functions of $|\bv|$ and $ \rho$. They can {\it not} depend of the {\it direction} of $\bv$ due to rotation invariance. 

We are not interested in fluctuations here, so we have dropped a noise term included by \cite{TT1, TT2, TT3, TT4, rean} in the velocity equation of motion. 

\vspace{.2in}

\section{The hydrodynamic theory of autochemotaxic flocks:  Equations of motion}

We now wish to develop a hydrodynamic theory of  autochemotaxic  flocking. This will be a continuum model which takes as its variables the three fields  introduced above:  the number density of boids $\rho(\br,t)$,   the boid velocity field $\bv(\br,t)$, and the concentration $\eta(\br,t)$ of the chemical signal (or chemo-attractant/repellent).

Let's begin with our modification of equation (\ref{KSC}) for the dynamics of the chemo-attractant,  $\eta$, in the Keller-Segel model . First, our model is not ``Malthusian"; that is, we will not consider birth and death of the boids. Therefore, we set $k_3=0$ in the equation of motion (\ref{KSrho}) for the boid density.

In addition, because of the coherent motion of the flock, we need to supplement the terms in the density equation of motion (\ref{KSrho}) with the convective $\nabla\cdot(\bv\rho)$ term in the Toner-Tu equation of motion (\ref{TTrho})  for the boid density. A similar term, but with a non-unit coefficient, can also appear in the chemo-attractant  equation of motion. Analogs of the ``convective derivative" terms with coefficients $\lambda_{1,2,3}$ that appear in the velocity equation of motion (\ref{TTv}) are allowed in the chemo-attractant  equation of motion as well. We also need to allow 
for {\it anisotropic} diffusive terms (i.e., anisotropic versions of the $k_{1,2}$ terms) in the density equation, and similar terms in the chemo-attractant  equation of motion, in much the way we introduced the anisotropic pressure $P_2$
in the velocity equation of motion (\ref{TTv})  in the Toner-Tu model.

In addition, all of the $k_i$'s can now depend on the local speed  $|\bv|$, as well as on $\rho$ and $\eta$.

The coupling of the chemo-attractant  concentration $\eta$ to the velocity dynamics must, by symmetry, be embodied entirely by making the pressures $P_{1,2}$, and  the phenomenological coefficients $U$, $D_{B,T,2}$, and $\lambda_{1,2,3}$, depend on $\eta$ as well as $|\bv|$ and $ \rho$. It is the dependence of the isotropic pressure on $\eta$ that leads to the instability we find, as we'll show below.

This reasoning leads to the following equations of motion for the boid density $\rho$, chemo-attractant  concentration $\eta$, and the velocity $\bv$\cite{TrueVelocity}: 

velocity

\bew
\beqn
\partial_t\bv + \lambda_1(\bv\cdot\nabla)\bv + \lambda_2(\nabla\cdot\bv)\bv + \lambda_3\nabla(|\bv|^2) &=& 
    U(|\bv|,\rho, \eta)\bv - \nabla P_1(|\bv|,\rho, \eta) - \bv(\bv\cdot\nabla P_2(|\bv|,\rho, \eta))
     \nn\\
     &+& D_B\nabla(\nabla\cdot\bv) + D_T\nabla^2\bv + D_2(\bv\cdot\nabla)^2 \bv \,, \label{KSTTvIntro}
     \eeqn
     \beqn
         \pp_t\rho&=& \nabla \cdot\bigg( k_1(|\bv|, \rho, \eta) \nabla \rho - k_2(|\bv|, \rho, \eta)\rho\nabla \eta +k_{1a}(|\bv|, \rho, \eta)\bv (\bv\cdot\nabla \rho) - k_{2a}(|\bv|, \rho, \eta)\rho\bv (\bv\cdot\nabla \eta)\bigg)-\nabla\cdot(\bv\rho)\,,\nn
    \label{KSTTrhoIntro}\\ 
    \eeqn
    \beqn
    \pp_t\eta&=& D_\eta\nabla^2\eta + k_6(|\bv|,\rho,\eta)
    -\lambda_{\eta1}(\bv\cdot\nabla)\eta - \lambda_{\eta2}(\nabla\cdot\bv)\eta { -\zeta_\eta\nabla\cdot(\bv \rho)}+D_{\eta\rho} \nabla^2\rho+D_{\eta v}\nabla^2|\bv|
    \nn\\
    &+&    D_{\eta a} \nabla\cdot[\bv(\bv\cdot\nabla)\eta]+D_{\eta\rho a} \nabla\cdot[\bv(\bv\cdot\nabla)\rho]+D_{\eta v a} \nabla\cdot[\bv(\bv\cdot\nabla)|\bv|],\label{KSTTetaIntro}
     \eeqn
     \ew
where we've defined 
\beq
k_6(|\bv|,\rho,\eta)\equiv
k_4(|\bv|, \rho, \eta) - k_5(\rho, \eta)\eta \,.
\label{k6def}
\eeq

As just discussed, we connect the concentration of the auto-attractant to the velocity equation by coupling them through the equations of state of the pressures $P_{1,2}$.

We have also introduced anisotropic analogs $k_{1a}$ and $k_{2a}$ of the  boid mobility $k_1$ and response to the chemotactant $k_2$, since we expect the response {\it along} the direction of the flock velocity $\bv$ to be different in general from the response perpendicular to that direction.

\section{Derivation of the Chemo-Attractant Instability} 

\subsection{The homogeneous polar fluid state}

We begin by identifying the steady state solution of our equations of motion that corresponds to a homogeneous moving flock. In a homogeneous state, all of our fields are constants, so we have 
\beqn
    \rho(\br,t) &= \rho_0 \,,\label{ss1}\\
    \eta(\br,t) &= \eta_0 \,,\label{ss2}\\
    \bv &= v_0\hat{x} \,.
    \label{ss3}
\eeqn
Note that the direction of $\bv$ is completely arbitrary, due to the rotation invariance of our model. We will henceforth choose our coordinate system so that the $x$-axis is the direction of the spontaneous velocity.

Inserting these constant ans\"{a}tze (\ref{ss1}), (\ref{ss2}) and (\ref{ss3}),  into our equations of motion  (\ref{KSTTvIntro}), (\ref{KSTTrhoIntro}) and (\ref{KSTTetaIntro}), it is clear that
all terms involving spatial or temporal derivatives vanish. It is easy to see that this implies that the density equation (\ref{KSTTrhoIntro}) is automatically satisfied. The 
chemo-attractant concentration equation in a steady, homogeneous state then reads
\beq
k_6(\rho_0, \eta_0) =0\,,\label{KSTTetahomo}
\eeq
while the velocity equation reduces to 
\beq
 U(v_0, \rho_0, \eta_0)=0 \,.
 \label{vcond}
 \eeq
 This amounts to two scalar algebraic equations with three unknowns. We clearly need one more condition. This can be obtained by fixing the number $N$ of boids contained in the volume $V$ of our system (in spatial  dimension $d=3$) (or the area in $d=2$).  Once we do so, we have, in a uniform system, fixed the density $\rho_0$ everywhere to be $\rho_0=N/V$ or $\rho_0=N/{\rm Area}$.  Inserting that value of $\rho_0$ into (\ref{KSTTetahomo}) then in principle determines $\eta_0$. (We will assume (\ref{KSTTetahomo}) has a unique solution for $\eta_0$ for any value of $\rho_0$.) 
 
 With these values of $\rho_0$ and $\eta_0$ in hand, we can then in principle use (\ref{vcond}) to determine the steady-state speed $v_0$. We will also assume that the solution of (\ref{vcond}) is unique, which it clearly will be if  $U(v_0, \rho_0, \eta_0)$ looks like figure  \ref{disspationgraph}.

\subsection{Linearization}
 
 Having found a steady state solution to our equations of motion that corresponds to a spatially homogeneous and uniformly moving flock, we now will determine the stability of this state. To do so, we begin by linearizing our equations of motion about our steady state solution. That is, we will write
 \beqn
    \rho(\br,t) &=& \rho_0 + \delta\rho(\br,t),\\
    \eta(\br,t) &= &\eta_0 + \delta \eta(\br,t),\\
    \bv(\br,t) &=& v_0\hat{x} + {\bf  \delta v}(\br,t) = (v_0 + \delta v_x(\br,t))\hat{x} + \bv_\perp(\br,t),\nn\\
\label{field_expand}
\eeqn
and then expand our equations of motion to linear order in the fluctuations $ \delta\rho(\br,t)$, $\delta \eta(\br,t)$, and ${\bf  \delta v}(\br,t)$ of the fields $\rho(\br,t)$, $\eta(\br,t)$, and  $\bv(\br,t)$. 

The process of linearization begins by expanding all of the $|\bv|$, $\rho$, and $\eta$ dependent parameters in the equations of motion to sufficiently high  order in  the fluctuations $ \delta\rho(\br,t)$, $\delta \eta(\br,t)$, and ${\bf  \delta v}(\br,t)$ of the fields $\rho(\br,t)$, $\eta(\br,t)$, and  $\bv(\br,t)$ to obtain all terms in the equation of motion to linear order in  $ \delta\rho(\br,t)$, $\delta \eta(\br,t)$, and ${\bf  \delta v}(\br,t)$. This gives us
\beqn
k_1 &=& D_{_C} \,,
\label{k1exp}\\
k_2 &=& D_{_C}\mu \,,
 \label{k2exp}\\
 k_3 &=& 0\,,
 \label{k3exp}\\
 k_6 &=& \gamma \delta\rho-{\delta\eta\over\tau}+\alpha\delta|\bv| \,,
 \label{k6exp}\\
 P_{1,2}(|\bv|, \rho, \eta)&=&{\rm constant}+\sigma_{1,2}\delta\rho-\kappa_{1,2}\delta \eta +  \nu_{1,2}\delta|\bv|\,,\nn\\
 \label{Pexp}\\
U(|\bv|,\rho, \eta) &=& -A\delta|\bv| + { {\bar B}}\delta\rho + C\delta \eta\,,
\label{Uexp}
\eeqn
 where $D_{_C}$, $\mu$,  $\gamma$,  $\tau$,  $\sigma_{1,2}$, $\kappa_{1,2}$,  $A$, ${ {\bar B}}$, and $C$ are all phenomenological constants. { We remind the reader that boids}  are not { being } born { and dying} ``on the wing''. { This means that the boid birth and death rate term $k_3$ is zero.} Most of the { other parameters in the model} can have either sign, but $\kappa_1$ and $A$  are both positive definite. The positivity of $A$ can be seen from figure  \ref{disspationgraph} , since $A$ is just $-1$ times the slope of $U(v)$ as it crosses the $v$ axis. The positivity of $\kappa_1$ means that the chemical signal {\it attracts} the boids ( i.e., that it {\it is} a chemo-{\it attractant}),  which is the case we wish to consider here. As we will see, it is this attraction that destabilizes the spatially uniform flocking state.
 
Note that we also have ${ {\bar B}}>0$ for systems in which the mean speed $v_0$ increases with increasing density, as it does for, e.g., the Vicsek model\cite{Vicsek}.

{ The parameters $\gamma$ and $\tau$ respectively set the sensitivity of the production rate and decay rate of the chemo-attractant.
}

Making the substitutions (\ref{field_expand}) and (\ref{k1exp}-\ref{Uexp}) in our equations of motion (\ref{KSTTvIntro}-\ref{KSTTetaIntro}), and keeping only terms to linear order in $\delta\rho$, $\delta\eta$, and $\delta\bv$,  we obtain our linearized equations of motion:

\bew
\beqn
\partialder{\delta\rho}{t} &=& D_{_C}\nabla^2\delta\rho - \mu D_{_C}(\rho_0\nabla^2\delta \eta) - \rho_0\partial_x\delta v_x - \rho_0\nabla_\perp\cdot\bv_\perp -  v_0\partial_x \delta\rho+D_{ca} v_0^2\pp_x^2\delta\rho-D_{ca} \mu_av_0^2\pp_x^2\delta\eta \,,  \label{rhoEq1} \\
\partialder{\delta \eta}{t} &=& D_\eta\nabla^2\delta\eta - \frac{\delta \eta}{\tau} + \gamma\delta\rho+\alpha \delta v_x{ -v_0\zeta_\eta\partial_x\delta\rho}-\lambda_{\eta1}v_0\pp_x\delta\eta-{ (\lambda_{\eta2}\eta_0 +\zeta_\eta\rho_0)}[\pp_x\delta v_x+\nabla_\perp\cdot\bv_\perp]\nn\\&+& D_{\eta\rho}\nabla^2\delta\rho+D_{\eta v}\nabla^2\delta v_x+ D_{\eta\rho a} v_0^2\pp_x^2\delta\rho+D_{\eta a} v_0^2\pp_x^2\delta\eta+D_{\eta va} v_0^2\pp_x^2\delta v_x \,,\label{NEq1}\\
\partialder{\delta v_x}{t} &=& v_0 (-A\delta v_x + { {\bar B}}\delta\rho + C\delta \eta)  -(\lambda_1+\lambda_2+2\lambda_3)v_0\pp_x\delta v_x-\lambda_2v_0(\nabla_\perp\cdot\bv_\perp)-(\sigma_1+v_0^2\sigma_2)\pp_x\delta\rho\nn\\&+&(\kappa_1+v_0^2\kappa_2)\pp_x\delta\eta-(\nu_1+v_0^2\nu_2)\pp_x\delta v_x+D_B[\pp_x(\nabla_\perp\cdot\bv_\perp)+\pp_x^2\delta v_x]+D_T\nabla^2\delta v_x+D_2v_0^2\pp_x^2\delta v_x \,,
\label{vxEq1}\\
 \nonumber \partialder{\bv_\perp}{t} &=& -\lambda_1v_0\pp_x\bvp-2\lambda_3v_0\nabla_\perp\delta v_x -\nabla_\perp (\sigma_1\delta\rho - \kappa_1\delta \eta +  \nu_1\delta v_x) 
+ D_B\nabla_\perp[\nabla_\perp\cdot\bv_\perp+\pp_x\delta v_x]\nn\\&&\,\,+ D_T\nabla^2_\perp\bv_\perp + D_{x}\partial_x^2\bv_\perp \,.\label{vperpEq1}
\eeqn
\ew
where $D_{x} \equiv D_T + D_2 v_0^2$.

We'll now use the equations to investigate the stability of the uniform  steady state.

\subsection{Demonstration that $\delta \eta$ and $\delta v_x$ are Fast Variables, and  their elimination}

We begin by observing that the fields $\delta v_x$ and $\delta\eta$ are ``fast" in the sense that their time derivatives do not vanish in the limit of extremely slowly spatially varying fields. In contrast, the fields $\rho$ and $\bvp$ are ``slow", or, to use another word, ``hydrodynamic": that is, their time derivatives {\it do} vanish in the limit of extremely slowly spatially varying fields.

As a result, if we are considering the most slowly evolving modes in the system at the longest wavelengths, the ``fast" fields $\delta v_x$ and $\delta\eta$ will, on the time scale of those slowest modes, relax very quickly back to values determined entirely by the slow fields. This means that those fast variables become effectively ``enslaved" to the hydrodynamic variables in the problem - in this case the boid density $\rho$. By ``enslaved", we mean that its value at any point in space and time is determined entirely by the instantaneous value of $\rho$ at the same point in space and time. It is this sort of ``enslavement" that enables hydrodynamic theories  in general  to eliminate ``fast" variables - that is, variables that do not relax infinitely slowly as the length scale under consideration goes to infinity- and work only with the ``slow" hydrodynamic variables.  

To see how this works in our problem, note that,  in the hydrodynamic limit of long wavelengths and large time scales, derivatives of the fluctuations of the fields are small compared to terms that contain the fluctuation with no derivatives. We will therefore expand the equations of motion for the fast fields $\eta$ and $\delta v_x$ in derivatives by writing
\beq
\delta\eta=\delta\eta_0+\delta\eta_1 \,\,\,\,\,\,, \,\,\,\,\,\, \delta v_x= \delta v_x^{(0)}+ \delta v_x^{(1)}\,,
\label{fastexpand}
\eeq
where
\beq
\delta\eta_0 \,, \delta v_x^{(0)}=O(\delta\rho) \,\,,\,\,
\delta\eta_1\,,  \delta v_x^{(1)}= O(\pp_t\delta\rho, \nabla\delta\rho, \nabla\cdot\bvp) \,.
\label{fastexpand2}
\eeq
{ Note that $\nabla\cdot v_\perp\ll\delta\rho$, as can be seen by the following argument:}
In order to satisfy Eq (IV.15), $\nabla\cdot v_\perp$ must be of order $(\pp_t\rho)$ or $
\pp_x\delta\rho$ (the only other term in (IV.15), namely  $\pp_x\delta v_x$, is clearly much 
smaller than $\nabla\cdot v_\perp$, since $x$-derivatives are much smaller than $\perp$ 
derivatives in the unstable regime (where $q_x\ll \bqp$), and $\delta v_x$ is $\ll\bvp$, since 
the former is massive). Therefore, $\nabla\cdot v_\perp$ is of order a {\it derivative} of $
\delta\rho$, and hence much less than $\delta\rho$, as we asserted. 

Therefore, to leading order in a hydrodynamic expansion, we can drop all terms with derivatives from the equations of motion
(\ref{NEq1}) and (\ref{vxEq1}). Dropping those terms, equations (\ref{NEq1}) and (\ref{vxEq1}) become: 

\beqn
0 &=& - \frac{\delta \eta_0}{\tau} + \gamma\delta\rho +\alpha \delta v_x^{(0)} \,,\label{NEq1lead}\\
0&=& v_0 (-A\delta v_x^{(0)} + { {\bar B}}\delta\rho + C\delta \eta_0) \,.
\label{vxEq1lead}
\eeqn
These equations can  trivially be solved to relate the ``fast" variables $\delta v_x$ and $\delta\eta$ to the ``slow" variable $\delta\rho$:
\beqn
\delta v_x^{(0)} &=&\left(\frac{{ {\bar B}} + \tau \gamma C}{A-\alpha\tau C}\right) \delta \rho=K_v\delta\rho,\,,
\label{rhovxelim}
\\
\delta \eta_0 &=&\left(\frac{(A\gamma+\alpha{ {\bar B}})\tau }{A-\alpha\tau C}\right) \delta \rho=K_\eta \delta \rho \,,
\label{rhoetaelim}
\eeqn
where we've defined
\beq
K_v\equiv
\frac{{ {\bar B}} + \tau \gamma C}{A-\alpha\tau C} \,\,\, \,\,\,, \,\,\, \,\,\, K_\eta\equiv\frac{(A\gamma+\alpha{ {\bar B}})\tau }{A-\alpha\tau C} \,.
\label{Kdefs1}
\eeq

The only important point to be noted about these rather complicated expressions is that both $K_v$ and $K_\eta$ are monotonically {\it increasing} functions of the 
chemo-attractant release rate $\gamma$,  provided $A$ and $C$ are positive. As pointed out earlier, $A$ definitely is positive.  Strictly speaking, $C$ could, in principle, have either sign. However, we expect intuitively that increasing the chemo-attractant  concentration should {\it increase} the speed of the boids (since most critters move faster when they smell food!). If this is the case, then $C$ will be positive as well. 

We'll show later that the fact that both  $K_v$ and $K_\eta$ are monotonically {\it increasing} functions of the chemo-attractant release rate $\gamma$ implies that the system will always eventually become unstable if we increase the chemo-attractant release rate $\gamma$ without bound.

{ The alert reader will notice that $K_v$ and $K_\eta$ both diverge as $A\to\alpha\tau C$ from above. The physics of this instability is simply that, when the denominator vanishes, the decay rate of one of the ``fast" modes vanishes. This can be seen by noting that the denominator in question is nothing but the determinant of the dynamical matrix for the fast fields if we set the slow field $\rho=0$. That determinant is simply the product of the two eigen-decay rates. While this instability would clearly be an interesting topic for future research,  we will not discuss it further here. Instead, we will focus our attention on the parameter regime $A>\alpha\tau C$, where this particular instability does not occur.}

Inserting (\ref{rhovxelim}) and (\ref{rhoetaelim}) into the linearized equation of motion (\ref{vxEq1}) for $\delta v_x$, and gathering terms proportional to one derivative (either spatial or temporal) of the hydrodynamic variables $\delta\rho$ and $\bvp$, gives
\beq
v_0(-A\delta v_x^{(1)}+C\delta\eta_1)=K_v\pp_t\delta\rho+G_v\pp_x\delta\rho+\lambda_2v_0\nabla\cdot\bvp \,,
\label{vx1st}
\eeq
where
\beqn
G_v&\equiv&\sigma_1+v_0^2\sigma_2
+K_v(\nu_1+\nu_2v_0^2+(\lambda_1+\lambda_2+2\lambda_3)v_0)\nn\\
&&\,-K_\eta(\kappa_1+v_0^2\kappa_2) \,.
\label{Gvdef}
\eeqn

Likewise, inserting (\ref{rhovxelim}) and (\ref{rhoetaelim}) into the linearized equation of motion (\ref{rhoEq1}) for $\delta \eta$, and gathering terms proportional to one derivative (either spatial or temporal) of the hydrodynamic variables $\delta\rho$ and $\bvp$, gives
\beq
 - \frac{\delta \eta_1}{\tau}+\alpha \delta v_x^{(1)}=K_\eta\pp_t\delta\rho+{ (\lambda_{\eta2}\eta_0+\zeta_\eta\rho_0)}\nabla\cdot\bvp+G_\eta\pp_x\delta\rho \,,
  \label{eta1st}
 \eeq
where 
\beq
G_\eta\equiv\lambda_{\eta1} v_0K_\eta+\lambda_{\eta2}\eta_0K_v { +v_0\zeta_\eta}\,.
\label{Getadef}
\eeq
Solving these two simple linear equations (\ref{vx1st}) and (\ref{eta1st}) for $\delta v_x^{(1)}$ and $\delta\eta_1$ gives
\beq
\delta v_x^{(1)}=-K_{v1}\pp_t\delta\rho-G_{v1}\pp_x\delta\rho-H_v\nabla\cdot\bvp
\label{dvx1sol}
\eeq
where
\beq
K_{v1}\equiv{v_0C\tau K_\eta+K_v\over v_0(A-\alpha C\tau)} \,,
\label{Kv1def}
\eeq

\beq
G_{v1}\equiv{G_v+C\tau v_0G_\eta\over v_0(A-\alpha C\tau)} \,,
\label{Kv1def}
\eeq

\beq
H_{v}\equiv{\lambda_2+{ (\lambda_{\eta2}\eta_0+\zeta_\eta\rho_0)}C\tau\over A-\alpha C\tau}\,,
\label{Hvdef}
\eeq
and
\beq
\delta\eta_1=-K_{\eta1}\pp_t\delta\rho-G_{\eta1}\pp_x\delta\rho-H_\eta\nabla\cdot\bvp \,,
\label{deltaetasol}
\eeq
where
\beq
K_{\eta1}\equiv{v_0A\tau K_\eta+\alpha\tau K_v\over v_0(A-\alpha C\tau)} \,,
\label{Keta1def}
\eeq
\beq
G_{\eta1}\equiv{\tau(\alpha G_v+A v_0G_\eta)\over v_0(A-\alpha C\tau)} \,,
\label{Geta1def}
\eeq
\beq
H_{\eta}\equiv{\tau(\alpha\lambda_2+{ (\lambda_{\eta2}\eta_0+\zeta_\eta\rho_0)}A)\over A-\alpha C\tau} \,.
\label{Hetadef}
\eeq

{ Where $K_{v1}$ and $K_{\eta 1}$ give the linear response of $\delta v_x^{(1)}$ and $\delta \eta_{1}$ to the time rate of change of the local density, $\delta \rho$. Likewise, the $G_{v1}$ and $G_{\eta 1}$ give the linear response to the spatial change in the direction of flock motion to the local density, $\delta \rho$. Finally, $H_\nu$ and $H_\eta$ give the linear response to the divergence of the local velocity that is perpendicular to the direction of flock motion, $\Delta\cdot\bv_\perp$.}

 Having eliminated the fast variables $\eta$ and $\delta v_x$ in favor of the slow variable $\delta\rho$, we can now write a set of closed equations of motion for  the density fluctuation $\delta\rho$ and perpendicular velocity field $\bvp$  by substituting equations (\ref{rhovxelim}), (\ref{rhoetaelim}), (\ref{dvx1sol}) and (\ref{deltaetasol}) into our equations of motion (\ref{rhoEq1}) and (\ref{vperpEq1}). This gives: 

\bew
\beqn
    \partialder{\delta\rho}{t} &= &D_{\rho\perp}\nabla_\perp^2\delta \rho +D_{\rho x}\pp_x^2\delta \rho- v_\rho\partial_x\delta\rho - \rho_0\nabla_\perp\cdot\bv_\perp+D_{\rho v}\pp_x(\nabla_\perp\cdot\bv_\perp)+\phi\pp_t\pp_x\delta\rho \,,  \label{rhoEq2} \\
     \partialder{\bv_\perp}{t} &=&- v_v\partial_x\bv_\perp -\left({B\over\rho_0}\right)\nabla_\perp \delta \rho 
    + D_{BR}\nabla_\perp(\nabla\cdot\bv_\perp) + D_T\nabla^2_\perp\bv_\perp + D_{x}\partial_x^2\bv_\perp+\nu_x\pp_x\nabla_\perp\delta\rho+\nu_t \pp_t\nabla_\perp\delta\rho \,, \nn\\
     \label{vperpEq2}
\eeqn
\ew
where we have defined: 
\beqn
    v_\rho &\equiv& \rho_0K_v + v_0   \,,\label{vrhodef}\\
    D_{\rho\perp} &\equiv& D_{_C}(1 - \mu K_\eta\rho_0) \,, \\
    D_{\rho x} &\equiv& D_{_C}+D_{Ca}v_0^2-D_{_C}\mu\rho_0K_\eta\nn\\&&\,\,-D_{Ca}\mu_av_0^2\rho_0K_\eta
    +\rho_0G_{v1} \,, 
\eeqn
\beqn
    D_{\rho v}&\equiv&\rho_0H_v\,,\\
    \phi&\equiv&\rho_0K_{v1}\,,\\
    v_v&\equiv&\lambda_1v_0\,,\\
    D_{BR}&\equiv&D_B+(2\lambda_3v_0+\nu_1)H_v+\kappa_1H_\eta\,,\\
    \nu_x&\equiv&\kappa_1G_{\eta1}+D_BK_v+(2\lambda_3v_0+\nu_1)G_v\,,\\
     \nu_t&\equiv&\kappa_1K_{\eta1}+(2\lambda_3v_0+\nu_1)K_{v1}\,,\label{nutdef}
  \eeqn
 and, most importantly, the ``inverse compressibility"
\beqn
    B&\equiv& \rho_0[\sigma_1 - K_\eta\kappa_1 +(2\lambda_3v_0+\nu_1)K_v]\nn\\
&=& \rho_0\bigg(\sigma_1 -\left[ \frac{\kappa_1(A\gamma+\alpha{ {\bar B}})\tau{ -}(2\lambda_3v_0+\nu_1)({ {\bar B}}+\gamma\tau C)}{A-\alpha\tau C}\right]\bigg)\,.\nn\\
\label{Bfinal}
\eeqn

This last expression is the most important result of this analysis, since it shows that autochemotaxis can make the inverse compressibility negative. One way this can be done is by increasing the parameter $\kappa_1$, which governs the sensitivity of the isotropic pressure to the chemo-attractant concentration. As can been by inspection of (\ref{Bfinal}), increasing $\kappa_1$ without bound while holding all other parameters fixed will inevitably make $B$ become negative, provided only that 
\beq
{(A\gamma+\alpha{ {\bar B}})\tau\over A-\alpha\tau C}>0 \,,
\label{kappa instab condition}
\eeq
which clearly is a very large portion of the parameter space of our hydrodynamic model.  

Likewise, increasing the chemo-attractant production rate $\gamma$ with all other parameters held fixed will also unavoidably drive the inverse compressibility $B$ negative, provided that 
\beq
{(\kappa_1A+(2\lambda_3v_0+\nu_1) C)\tau\over A-\alpha\tau C}>0 \,.
\label{tau instab cond}
\eeq

Finally, increasing the chemo-attractant lifetime $\tau$ with all other parameters held fixed will also certainly  drive the inverse compressibility $B$ negative if, for example, all of the parameters $\alpha$, ${ {\bar B}}$, $C$, and the combination $2\lambda_3v_0+\nu_1$ are all positive. Indeed, in this case, $B$ is guaranteed to  turn negative as $\tau$ is increased from zero for some $\tau<{A\over\alpha C}$.

The forms of the equations of motion (\ref{rhoEq2}) and (\ref{vperpEq2}) are {\it identical} to the linearized equations of motion obtained in \cite{rean} for stable flocks without chemotaxis. This was inevitable,  since the symmetries and conservation laws of our model are the same as those of an ordinary, stable flock. The only long-wavelength effect of the autochemotaxis is to provide a mechanism to drive the inverse compressibility $B$ negative, as discussed above. { Any attractive interaction, which is sufficiently strong to drive the inverse compressibility negative, will obey the same hydrodynamic description put forward in this paper.} In the next section, we will show that when $B$ does become negative, the homogeneous phase of the flock becomes unstable.

\vspace{.2in}

\subsection{Mode Structure}

To investigate the mode structure of our linearized model, we proceed exactly as in \cite{TT1,TT2,TT3,TT4, rean},  by seeking solutions of the linearized equations of motion  (\ref{rhoEq2}) and (\ref{vperpEq2})  for plane-wave normal modes, in which all fields are proportional to $\exp{[i\boldsymbol{q}\cdot \boldsymbol{r}-i\omega t]}$. Doing so, the equations of motion then read:
\bew
\begin{align}
    \bigg[-i(\omega-v_\rho q_x) +  D_{\rho\perp}q_\perp^2 +D_{\rho x}q_x^2  -\phi\omega q_x \bigg] \delta\rho &= -i\rho_0\bq_\perp\cdot\bv_\perp -  D_{\rho v}q_x(\bq_\perp\cdot\bv_\perp) 
    \label{rhoEq3} \,,\\
    \bigg[-i\omega + iv_vq_x + D_Tq^2_\perp + D_{x}q^2_x \bigg]\bv_\perp + D_{BR}\bq_\perp(\bq_\perp\cdot \bv_\perp) &= \bigg[-i\frac{B}{\rho_0}+\nu_t\omega-\nu_xq_x\bigg]\bq_\perp\delta\rho \,. \label{vperpEq3}
\end{align}
\ew
We decouple these by projecting (\ref{vperpEq3}) perpendicular to and along $\bq_\perp$. That is, we write 
\beq
\bv_\perp = v_L\hat{q}_\perp + \bv_T \,,
\label{v decomp}
\eeq 
with the ``transverse"  components $\bv_T$, by definition, perpendicular to $\bqp$ (that is, $\bv_T\cdot\hat{q}_\perp = 0$), while the single ``longitudinal" component $v_L = \frac{\bq_\perp\cdot\bv_\perp}{|q_\perp|}$ is the projection of $\bvp$ onto $\bqp$.

This split decouples $\delta\rho$  and $v_L$ from the $d-2$  transverse modes $\bv_T$, as can be seen by  inserting the decomposition (\ref{v decomp})  into the $\rho$ equation of motion (\ref{rhoEq3}), and projecting the velocity equation of motion (\ref{vperpEq3}) in the transverse and longitudinal directions  (i.e., along and perpendicular to $\bqp$).  This gives:
\bew
\begin{align}
    \bigg[-i(\omega - v_\rho q_x) + \Gamma_\rho(\bq) -\phi\omega q_x \bigg] \delta\rho &= -i\rho_0q_\perp v_L  -  D_{\rho v}q_xq_\perp v_L, \label{rhoEq4} \\
    \bigg[-i(\omega - v_vq_x) + \Gamma_L(\bq)\bigg] v_L &= \bigg[-i\frac{B}{\rho_0}+\nu_t\omega-\nu_xq_x\bigg]q_\perp \delta \rho , \label{vLEq1} \\
    \bigg[-i(\omega - v_vq_x) + \Gamma_T(\bq)\bigg] \bv_T &= {\bf 0} \,. \label{vTEq1} 
\end{align}
\ew
Here we've defined $\Gamma_T  = D_T q^2_\perp + D_{x}q_x^2$, $\Gamma_L  = D_{_{L\perp}} q^2_\perp + D_{x}q_x^2$, $D_{_{L\perp}} = D_T + D_{BR}$, and $\Gamma_\rho =  D_{\rho\perp}q_\perp^2+D_{\rho x}q_x^2$.

Altogether we have $d$ frequencies to determine. Since the transverse modes are decoupled, we can immediately write down the transverse dispersion relation implied by equation (\ref{vTEq1}):

\begin{align}
    \omega = -i\Gamma_T + v_vq_x
\end{align}
All these $d-2$ modes decay, provided $D_Tq_\perp^2 + D_{x}q_x^2 > 0$.  Thus, there is no instability in these modes, even when the inverse compressibility $B$ becomes negative (unsurprisingly, since
$B$ does not enter the transverse equations of motion).

The instability arises from the coupled density $\rho$ and longitudinal velocity $v_L$ modes, as can be seen from equations (\ref{rhoEq4}) and (\ref{vLEq1}), which can be rewritten: 

\bew
\beqn
-i\omega\delta\rho&= &-D_{\rho\perp}q_\perp^2\delta \rho -D_{\rho x}q_x^2\delta \rho- iv_\rho q_x\delta\rho - i\rho_0q_\perp v_L-D_{\rho v}q_xq_\perp v_L+\phi\omega q_x\delta\rho \,,  \label{rhoEq4.2} \\
     -i\omega v_L &=&-i v_vq_x v_L -i\left({B\over\rho_0}\right)q_\perp \delta \rho 
    - D_{_{L\perp}}q^2_\perp v_L - D_{x}q^2_x v_L -\nu_xq_xq_\perp\delta\rho+\nu_t \omega q_\perp\delta\rho \,\,.  \label{vperpEq4.2}
\eeqn
\ew
As shown in \cite{TT1, TT2, TT3,TT4, rean}, the eigenfrequencies of these equations, to linear order in $q$, are
\begin{align}
    \omega_\pm = \frac{1}{2}\bigg[q_xv_\rho  + q_xv_v \pm \Delta \bigg],\label{omegaqQ} \\
    \Delta =\sqrt{(v_v - v_\rho )^2q_x^2 + 4Bq_\perp^2}.
    \label{omegaq}
\end{align}

 In contrast to prior work, however, we now know that autochemotaxis can drive the inverse compressibility $B$ negative. When this happens, it is clear from 
(\ref{omegaq}) that modes whose wavevectors $\bq$ lie in the region in which
\begin{align}
    (v_v-v_\rho )^2 q_x^2 + 4Bq_\perp^2< 0 \,, \label{instabilityCondition}
\end{align}
will be unstable, thereby destabilizing the spatially uniform flocking state. This implies the instability lies in the region 
\beq
|q_x|<{2\sqrt{|B|}\over|v_v-v_\rho |}|\bqp|=\epsilon|\bqp|  \,,
\label{IS2}
\eeq
where we've defined:

\beq
\epsilon\equiv{2\sqrt{|B|}\over|v_v-v_\rho|} \,.
\label{epsdef}
\eeq
(Recall that in the unstable region $B<0$).
If we are near the instability on the unstable side, $|B|\ll(v_v-v_\rho )^2$, and so (\ref{IS2}) implies $\epsilon<<1$, and so the unstable wavevectors satisfy $|q_x|\ll|\bqp|$, as claimed in the introduction. 

The window (\ref{IS2}) gives the region in which, at sufficiently small $\bq$, the flock becomes unstable with a growth rate that is {\it linear} in $q\equiv|\bq|$. Small $\bq$'s that lie outside this region, however, are not necessarily stable. This is because, for $|q_x|>\epsilon|\bqp|$, the eigenfrequencies $\omega$ {\it to linear order in} $q$, 
are real, as can be seen by inspection of equation \ref{omegaqQ}. Since the stability or instability is determined entirely by the {\it imaginary} part of $\omega$, we must therefore, for $\bq$'s in the regime $|q_x|>\epsilon|\bqp|$,  work to higher (in fact, second) order in $\bq$ in order to determine the absolute instability limit. This proves, as we shall see, to be have a larger opening angle at small $\bq$ than the limit (\ref{IS2}), but the new limit has the same $\sqrt{|B|}$ scaling with $B$. In contrast to equation (\ref{IS2}), however, the instability is only limited to a narrow window if $|v_v-v_\rho|$ 
exceeds some finite threshold. See equation (\ref{inst bound}) for more details.

It is instructive to rewrite the dispersion relation (\ref{omegaq}) in terms of two  complex, direction dependent sound speeds $c_\pm(\theta_\bq)$:
\beq
\omega=c_\pm(\theta_\bq)q \,,
\label{cdef}
\eeq
where $q \equiv|\bq|$, and $\theta_\bq$ is the angle between the $\bq$ and the $\hat{x}$ direction (i.e., the direction of mean flock motion  $\vb$). Using  $q_x = q \cos{\theta}$ and $q_\perp = q\sin{\theta}$ in (\ref{omegaq}), we obtain the  expression of references \cite{TT1,TT2,TT3,TT4,rean} for 
$c_\pm(\theta_\bq)$:
\bew
\begin{align}
    c_\pm(\theta_\bq) = \frac{1}{2}\bigg[(v_\rho  + v_v)\cos(\theta_\bq) \pm \sqrt{(v_v - v_\rho )^2\cos^2(\theta_\bq) + 4B\sin^2(\theta_\bq)}\bigg].
    \label{cpm}
\end{align}
\ew
from this expression, we can see that $\omega_+$ acquires a positive imaginary part, signaling instability, for directions of propagation $\theta_\bq$ near $\pm\pi/2$; i.e., perpendicular to the direction of mean flock motion. 
Specifically, 
the instability occurs for
\beq
|\theta\pm\pi/2|<{2\sqrt{|B|}\over|v_v-v_\rho|} \,.
\label{instab angle}
\eeq

For directions of propagation $\theta_\bq$ between $0$ and ${\pi\over2}$, this condition is equivalent to
\beq
\theta_\bq>\theta_c\equiv{\pi\over2}-{2\sqrt{|B|}\over|v_v-v_\rho|} \,.
\label{thetacdef}
\eeq

This result holds only for very small $q$.  As noted earlier, even at very small $q$, it only gives the region in which the instability growth rate is {\it linear} in $q$; the possibility of an $O(q^2)$ instability outside this small angular window still exists. Furthermore,  at larger $\bq$, higher order terms in the equation of motion become important. As a result, even $\bq$'s whose directions lie in the narrow angular wedge (\ref{instab angle}) do not remain unstable out to arbitrarily large $|\bq|$.

To  investigate both of these possibilities, we'll now go beyond leading order in $q$ for systems close to the instability threshold. Assuming that we continue to have $q_x\sim\epsilon|\bqp|$ with $\epsilon\ll1$, we can greatly simplify the eigenvalue conditions (\ref{rhoEq2}) and (\ref{vperpEq2}). We'll also assume, and verify {\it a posteriori}, that the typical unstable wavevector satisfies { $|\bqp|\propto{\epsilon}$}. Then the various terms in (\ref{rhoEq4.2}) scale as follows: 
\beqn
&i\omega \drho \sim \epsilon^2 \drho \,\,\,\,, \,\,\,\,D_{\rho \perp} q_{\perp}^2 \drho \sim \epsilon^2 \drho  \,, \nonumber\\ 
&D_{\rho x} q_x^2 \drho \sim \epsilon^4 \drho  \,\,\,\,, \,\,\,\, iv_\rho q_x \drho \sim \epsilon^2 \drho \,, \nonumber\\ 
&i\rho_0q_\perp v_L \sim \epsilon v_L  \,\,\,\,, \,\,\,\, D_{\rho v} q_xq_\perp v_L \sim \epsilon^{3} v_L \,,\nonumber\\ 
&\phi \omega q_x \drho \sim \epsilon^4 \drho \, \nonumber.
\eeqn
Thus, as we approach the transition, so that $\epsilon\to0$, we can neglect the terms
$D_{\rho x}q_x^2\delta \rho$ and $\phi\omega q_x\delta\rho$ relative to $\omega\rho$, and also drop the $D_{\rho v}q_xq_\perp v_L$ term relative to the $v_\rho q_x\delta\rho$ term. Doing so reduces our $\delta\rho$ equation of motion  (\ref{rhoEq4.2}) to the far more manageable form:
\begin{align}
   [-i(\omega-v_\rho q_x) + D_{\rho \perp}q_{\perp}^2]\delta\rho + i \rho_0q_\perp v_L &=0  \,.\label{rhoSimp}
\end{align}

Performing this sort of power counting on the $v_L$ equation of motion (\ref{vperpEq4.2}) gives:
\begin{align*}
    i \omega v_L &\sim \epsilon^2 v_L \,\,\,\,, \,\,\,\, \,\,\,\, \,\,\,\, \,\,\,\, \,\,
    v_v q_{x} v_L \sim \epsilon^2 v_L\,, \\
    B q_\perp \drho &\sim \epsilon^{3} \drho \,\,\,\,, \,\,\,\,
    D_{_{L\perp}} q_{\perp}^2 v_L \sim \epsilon^2 v_L \,,\\
    D_{x} q_x^2 v_L &\sim \epsilon^4 v_L \,\,\,\,, \,\,\,\,
    \nu_x q_xq_\perp \drho \sim \epsilon^{3} \drho\,, \\
    \nu_t \omega q_\perp \drho &\sim \epsilon^{3} \drho \,, 
\end{align*}
which shows that we can drop the $D_{x}q_x^2$ term in the $v_L$ equation of motion (\ref{vperpEq4.2})  relative to the $D_{_{L\perp}} q_\perp^2$ term. Doing so gives:

\bew
\begin{align}
    [-i(\omega-v_vq_x)+D_{_{L\perp}}q_\perp^2]v_L + \bigg[-i\frac{|B|}{\rho_0}q_\perp + \nu_xq_xq_\perp -\nu_t\omega q_\perp\bigg]\delta\rho &= 0 \,.
    \label{vLsimp}
\end{align}
\ew
These simplified equations of motion (\ref{rhoSimp}) and (\ref{vLsimp}) can be combined into a single vector equation:
\beq
\bm{M}(\omega) \left[
\begin{array}{cc}
\delta\rho\\[10pt]
v_L
\end{array}
\right]=\left[
\begin{array}{cc}
0\\[10pt]
0
\end{array}
\right] \,,
\label{vectorEOM}
\eeq
where the matrix $\bm{M}$ is given by:
\bew
\[
\bm{M} = \left(
\begin{array}{cc}
  [-i(\omega-v_\rho q_x) + D_{\rho \perp}q_{\perp}^2] & i \rho_0q_\perp \\[10pt]
\bigg[-i\frac{|B|}{\rho_0}q_\perp + \nu_xq_xq_\perp -\nu_t\omega q_\perp\bigg] &  [-i(\omega-v_vq_x)+D_{_{L\perp}}q_\perp^2]
\end{array}
\right)\;.
\]
\ew
The eigenvalue condition that determines the eigenfrequencies $\omega$ is clearly simply
\beq
\det\bm{M}(\omega)=0 \,,
\label{EVcond}
\eeq
which leads to

\begin{figure}
        \centering
       \includegraphics[width=\linewidth]{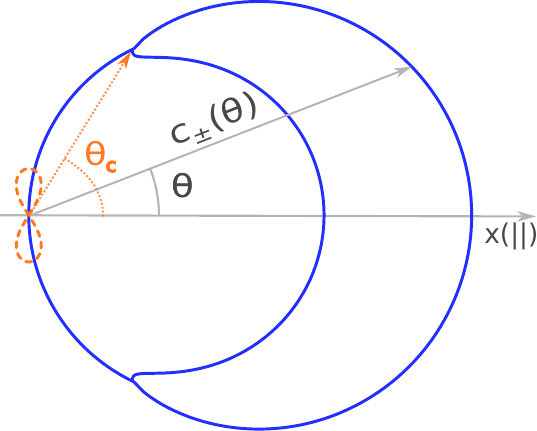}
        \caption{Polar plot of the direction dependent sound speed $c_\pm(\theta)$ from equation (\ref{cpm}) ($B=-0.025,~v_\rho =1.50, ~v_v=1$  (all units arbitrary).). The horizontal axis, $x(\parallel)$, is the mean direction of the flock's motion. The solid blue line is the real part of the sound speed. The dashed orange line is the imaginary part of the sound speed. Wavevectors with non-zero imaginary speed values are unstable. This instability region is highly anisotropic,  with all of the unstable wavevectors very nearly perpendicular to $y(\parallel)$.  Note that the bifurcation of the real part coincides with the opening angle of the imaginary part{ ,
        $\theta_c$}.}
        \label{SoundSpeedPlot}
\end{figure}

\bew
\begin{align}
    \label{eigenvalueCondition}
    -\omega^2 + \omega[(v_v+v_\rho)q_x &- i(D_{L\perp} + D_{\rho\perp} - \rho_0\nu_t)q_\perp^2] \nonumber \\
    & - |B| q_\perp^2 + i(v_\rho D_{L\perp} + v_vD_{\rho \perp}  -\rho_0\nu_x)q_xq_\perp^2 + D_{L\perp}D_{\rho\perp} q_\perp^4 - v_\rho v_v q_x^2 = 0 \,.
\end{align}
\ew

 \begin{figure}
        \centering
      \includegraphics[width=0.5\linewidth]{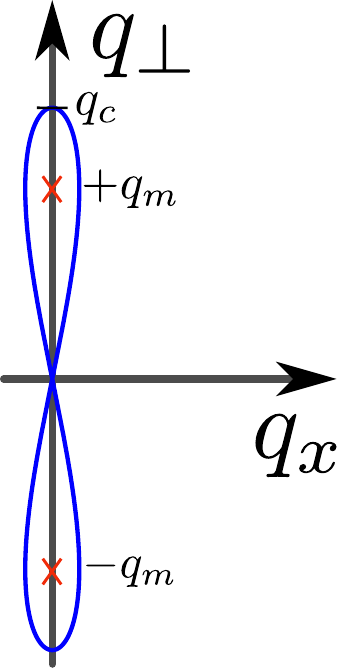}
        \caption{A plot of the highly anisotropic instability region in $q$-space.  There are unstable modes at all $\bq$'s inside the two loops of this lemniscate. The largest unstable wavevector $q_c$ is marked with a black dash at the highest point on the  instability boundary. The wavevectors $q_m$ with the fastest growing modes are identified with red crosses. }
        \label{window}
\end{figure}

We'll first determine the instability boundary.
Instability occurs for those $\bq$'s for which the imaginary part  ${\rm Im}\,\omega(\bq)$ of $\omega(\bq)$ is positive ; modes with  ${\rm Im}\,\omega(\bq)$ negative are stable. Therefore, the boundary between stable and unstable regions of $\bq$ must be the locus on which ${\rm Im}\,\omega(\bq)=0$; i.e., the locus on which $\omega$ is purely real. We can determine this locus by first taking the imaginary part of (\ref{eigenvalueCondition}) when ${\rm Im}\,\omega(\bq)=0$. This implies 
\beq
-(D_{_{L\perp}}+D_{\rho \perp}-\nu_t\rho_0)q_\perp^2\omega+(v_vD_{\rho \perp}+v_\rho  D_{_{L\perp}} -\rho_0\nu_x) q_\perp^2q_x=0 \,,
\label{imocond}
\eeq
which can be trivially solved for the value $\omega_{inst}$ of  $\omega$  at which the instability occurs:
\beq
\omega_{inst}=\left({v_vD_{\rho \perp}+v_\rho  D_{_{L\perp}} -\rho_0\nu_x\over D_{_{L\perp}}+D_{\rho \perp}-\rho_0\nu_t}\right)q_x \,.
\label{omegabound}
\eeq

Plugging this value of $\omega$ into the {\it real} part of 
 (\ref{eigenvalueCondition}) then leads to a linear equation for $q_x^2$, whose solution is

 \begin{align}
     q_x^2 = \frac{|B|q_\perp^2-D_{\rho\perp}D_{_{L\perp}}q_\perp^4}{ V^2} \,,
     \label{inst bound}
 \end{align}
 where we've defined

\beqn
    D_3 &\equiv&  D_{_{L\perp}} + D_{\rho\perp}-\rho_0\nu_t \,,\\
    D_6 &\equiv&  {4\over v_4}(v_\rho D_{_{L\perp}} + v_vD_{\rho\perp} - \nu_x\rho_0) - 2D_3\,,\\
     V^2 & \equiv& \left[{(v_v - v_\rho)^2\over4}-{D_6^2(v_v+v_\rho)^2\over16D_3^2}\right] \,.
    \label{Ddefs}
\eeqn

The alert reader will recognize (\ref{inst bound})
as the shape of the  stability boundary in wavevector space, as depicted in figure \ref{q-unstable} , which we present again here,  with a new caption, as figure \ref{window}. She will also recognize this as equation (4) of the associated short paper\cite{ASP}.

\begin{figure}
        \centering
      \includegraphics[width=\linewidth]{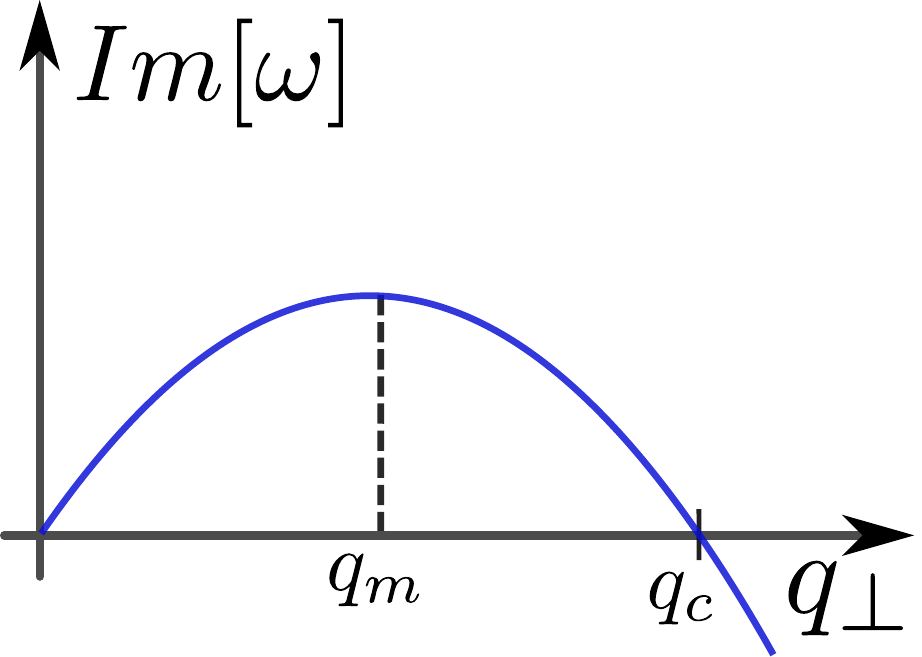}
        \caption{Plot of the growth rate (\ref{growth rate}). All wave vectors beyond $q_c$ decay. The fastest growth occurs at wavevector $q_\perp=q_m$, the position of the maximum of this curve.}
        \label{instabilityThreshold}
\end{figure}

Note that the maximum value $q_c$ of $|\bqp|$ in the unstable region is the value 
\beq
q_c=\sqrt{|B|\over D_{_{L\perp}}D_{\rho\perp}}
\label{qcdef}
\eeq
of $|\bqp|$ at which the right hand side of (\ref{inst bound}) vanishes. { This is the justification for the preceding scaling argument, which assumed that $|q_\perp| \lesssim\epsilon$.}

Having found the instability boundary, we'd now like to determine the wavevector of the fastest growing mode. As a first step, we'll solve (\ref{eigenvalueCondition}) for $\omega$:

\begin{align}
    \omega &= \frac{1}{2}\bigg[v_4q_x - iD_3q_\perp^2 \pm \sqrt{S}\bigg] \label{dispersionRelation77}\\
    S &=(v_\rho-v_v)^2q_x^2 - D_5^2 q_\perp^4 - 4 |B|q_\perp^2 + i v_4D_6q_xq_\perp^2 
\end{align}
where we've defined
\begin{align}
    v_4 &\equiv v_\rho+v_v,\\
    D_5^2 &\equiv  D_3^2 -4D_{_{L\perp}}D_{\rho\perp}.\label{dfive}
\end{align}
Let's first maximize the imaginary part of this over $q_x$. Note that all of the $q_x$ dependence of the imaginary part comes from the square root. Therefore, we need only maximize the imaginary part of that square root. This implies we want to find the value of $q_x$ at which ${\pp{\rm Im}\sqrt{S}\over\pp q_x}=0$. This condition is equivalent to saying that ${\pp \sqrt{S}\over\pp q_x}={1\over2\sqrt{S}}\left(\pp S\over\pp {q_x}\right)$ is real. Squaring both sides of this condition implies that ${1\over S}\left(\pp S\over\pp {q_x}\right)^2$ is also real.  Hence, at the wavevector of the fastest growing mode,
\begin{align}
   {\left(\pp a\over\pp_{q_x}\right)^2-\left(\pp b\over\pp_{q_x}\right)^2+2i\left(\pp a\over\pp_{q_x}\right)\left(\pp b\over\pp_{q_x}\right)\over S}={\rm real}  \,,
   \label{MG4.1}
\end{align}
where $a\equiv(v_\rho-v_v)^2q_x^2  - D_5^2 q_\perp^4 - 4 |B|q_\perp^2$ and $b\equiv  D_6q_xq_\perp^2$ are the real and imaginary parts of $S$ respectively.
Calculating the required partial derivatives and performing this algebra, we can rewrite this condition as: 
\bew
\beq
4(v_\rho-v_v)^2\left({(v_\rho-v_v)^2q_x^2 - \left({D_6^2\over4(v_\rho-v_v)^2}\right) q_\perp^4  + i D_6q_xq_\perp^2\over(v_\rho-v_v)^2q_x^2 -D_5^2 q_\perp^4 - 4 |B|q_\perp^2 + i D_6q_xq_\perp^2}\right)={\rm real}
\label{max cond}
\eeq
\ew

Note that the imaginary parts of the numerator and denominator{ of  the parenthetical term in} (\ref{max cond}) are equal. Therefore
the condition can only be satisfied in two ways: either

\noindent 1) the real parts of the numerator and denominator are also equal, or

\noindent 2) the imaginary parts of the numerator and denominator both vanish.

Note that condition 2) implies $q_x=0$. (We can't have $q_\perp=0$ and $q_x\ne 0$, since $q_x\ll q_\perp$ in the unstable region.) This proves to be the condition at the maximum.

We will prove that condition 2) must hold by showing that condition 1) is impossible.  Assuming condition 1) is satisfied leads to the following constraint on $q_\perp$:

\begin{align}
\bigg( -D_5^2 + \frac{D_6^2}{4(v_v-v_\rho)^2}\bigg)q_\perp^2 &=4|B|.\label{MG51}
\end{align}
This condition can never be satisfied in the unstable region, at least for small $\nu_{x,t}$. To see this, note that when $\nu_x$ and $\nu_t$ are both zero, the coefficient of $q_\perp^2$ on the left hand side of (\ref{MG51}) vanishes, as can be seen from the expressions  (\ref{dfive}) and (\ref{Ddefs}) for $D_{5,6}$ evaluated for $\nu_{x,t}=0$. This means that for small $\nu_x$ and $\nu_t$, the
the coefficient of $q_\perp^2$ on the left hand side of (\ref{MG51}) is guaranteed to be small compared to $D_{_{L\perp}}D_{\rho\perp}$.  That in turn implies that $|\bqp|>>\sqrt{|B|\over D_{_{L\perp}}D_{\rho\perp}}=q_c$. However,  such wavevectors lie outside the unstable region, in which $|\bqp|<q_c$. Hence, this solution is not acceptable, at least for small  $\nu_x$ and $\nu_t$.  This is the limit on which we will henceforth focus. In this limit, we must therefore have $q_x=0$ at the wavevector of fastest growth.

Setting $q_x=0$ in our expression (\ref{dispersionRelation77}) for $\omega$ greatly simplifies that expression. Taking ${\rm Im}~\omega$ gives the growth rate

\begin{align}
    { {\rm Im}~\omega} = -D_3q_\perp^2 + \sqrt{4|B| q_\perp^2 + D_5^2q_\perp^4}
    \label{growth rate}.
\end{align}
Maximizing this over $q_\perp$ gives, after a little algebra, a quadratic equation for the square of the value $q_m$ of $q_\perp$ at which the growth rate is maximized:
\beq
\left({D_5^2\over D_{_{L\perp}}D_{\rho\perp}}\right)q_m^4+4q_c^2q_m^2-q_c^4=0 \,,
\label{qmcond}
\eeq
 whose only real solutions are  
\beq
q_m=\pm q_c\Bigg( \frac{1}{D_5}\sqrt{D_{_{L\perp}}D_{\rho\perp}\bigg[\sqrt{4+\left({D_5^2\over D_{_{L\perp}}D_{\rho\perp}}\right)}-2\bigg]}\Bigg) \,,
\label{qm}
\eeq 
whose magnitude the reader can verify for herself is always less than $q_c$, as it must be.

\subsection{Intuitive explanation for the anisotropy of the instability}{\label{intuit}}

Why is the instability we just found restricted to modes with their wavevectors $\bq$ almost perpendicular to
 the mean velocity  $\vb$? 
 The explanation  is the competition between convection and pressure forces.

 By convection, we mean the $v_\rho$ and $v_v$ terms in equations (\ref{rhoEq2}) and (\ref{vperpEq2}) respectively. These would, in the absence of other effects, cause any instantaneous local configuration of $\rho(\br, t)$ or $\bv(\br,t)$
 to simply be transported along the direction of flock motion at speeds   $v_\rho$ and $v_v$ respectively. Naively, one might have expected  $v_\rho$ and $v_v$ to both simply equal the speed of the flock $|\vb|$ itself, since this would simply correspond to the fluctuations moving along with the flock as it moves. However, as explained in reference \cite{TT1}, due to the lack of Galilean invariance in our system, there is no reason that either of these speeds should be equal to $|\vb|$, since the flock itself does not provide the only observable Galilean reference frame in the problem; there is also the frictional substrate. Nor is there any reason that $v_\rho$ and $v_v$ should be equal to each other. Indeed, their {\it inequality} plays a crucial role in limiting the instability, as can be seen from, e.g., the expression (\ref{epsdef}) for $\epsilon$, which shows that $\epsilon\to\infty$ if $v_\rho=v_v$.  That is, the window of instability would have an opening angle that covers all space, making all wavevectors unstable.
 
\begin{figure*}
    \centering
    \includegraphics[width=\linewidth]{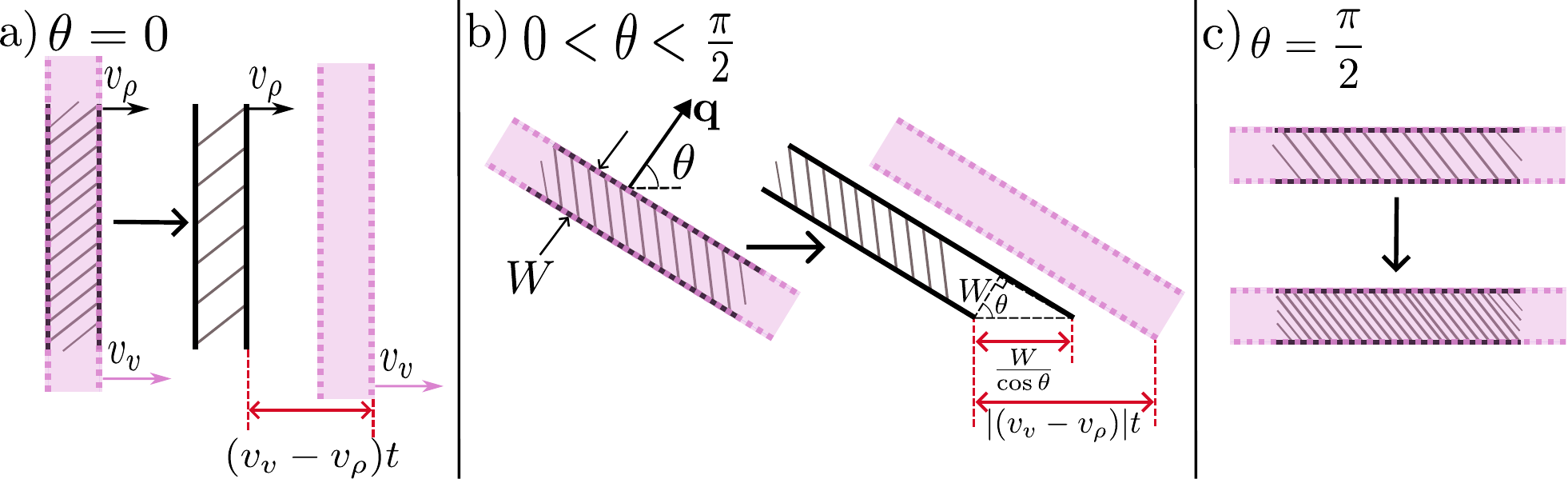}
    \caption{
     Three density and velocity fluctuations are depicted at different angles $\theta$ relative to the mean motion of the flock. The density fluctuation is depicted as a short rectangular region with cross hatching. The density  of the hash marks indicates the density in that region. The velocity fluctuation is depicted as a long pink rectangular region with dashed edges. Both regions propagate to the right, which is the in mean direction of motion of the flock in all three panels, but at different speeds $v_\rho$ and $v_v$ respectively,  and extend infinitely in the directions parallel to their depicted boundary.  a) $\theta = 0$, $c_+=v_v>v_\rho=c_-$. The velocity fluctuation separates from the density fluctuation. The density fluctuation begins to decay by diffusion, and the system is stable against such fluctuations. b) $0<\theta <{\pi\over2}$,the velocity and density bands still separate over time, but by an a
 distance that's reduced by a factor of $\cos(\theta)$ when projected along the normal to the bands. As a result, if $\theta>\theta_c$, the bands do not fully separate before the instability sets in, and those fluctuations therefore grow, destablizing the system. c) $\theta = \pi/2$. The density and velocity fluctuations both propagate parallel to the band itself, so they continue to overlap for all time. This is therefore the {\it most} unstable direction.
 }
    \label{allbands}
\end{figure*}

 To understand why the inequality of $v_\rho$ and $v_v$ is so crucial, consider the evolution of a small fluctuation in which both the  density and velocity depart from their mean values in a band perpendicular to the direction of mean flock motion, as illustrated in figure \ref{allbands}. This corresponds to a fluctuation with wavevectors making an angle $\theta_\bq=0$ with the mean velocity $\vb$. The fluctuation in the density creates a pressure gradient, which, because the inverse compressibility $B$ is negative, causes the velocity field to increase in the direction that tends to {\it increase} the fluctuation in the density. That is, if the density is {\it higher} in our band, the pressure forces tend to increase the velocity {\it into} the band, making the density grow even more. The opposite will happen if the density is lower than average in our band: pressure forces will tend to make the velocity flow {\it out}, carrying boids out of the band and reducing the density even further.
 
 It is this positive feedback mechanism that is responsible for the familiar instability of negative compressibility equilibrium systems. In {\it our} system, however, this mechanism must compete against the aforementioned convection. For the band direction just discussed, convection stabilizes the system by separating the band of perturbed velocity from the band of perturbed density, because they propagate at different speeds. This is illustrated { in figure \ref{allbands}a}. 

Once the velocity fluctuation is separated from the density fluctuation, both simply decay away by diffusion. Therefore, this orientation of the bands is stable.

Now consider the other extreme case, in which the band runs {\it parallel} to the direction of mean flock motion, as illustrated { in figure \ref{allbands}c}. Now the relative displacement of the velocity and density fluctuations makes no difference: the two bands of each still overlap, since the bands are infinitely extended in the direction of that relative motion. Hence, the velocity fluctuation continues to sit right on top of the density fluctuation, allowing the pressure induced instability to occur.

We can even derive the critical instability angle $\theta_c$ by this argument.  To see this, consider a fluctuation in which the initial band now runs at an angle $\theta$ to the vertical, (which corresponds to the component wavevectors $\bq$ making an angle $\theta$ with the horizontal (i.e., with the $x$-axis)), as illustrated { in   \ref{allbands}b}. Now the bands will again separate by a distance $|v_v-v_\rho|t$ from each other in a time $t$, but that separation {\it is in the $x$-direction}. The {\it projection} of that displacement {\it perpendicular}  to the bands (which is what determines when the bands actually separate, of course) is therefore only $|v_v-v_\rho|t\cos\theta|$. 

How long do these bands have to separate before the instability sets in? Well, roughly speaking, the instability growth rate is $\sqrt{|B|}q$, where the wavenumber $q\sim1/W$, with $W$ the width of the band. Hence, the characteristic time for the instability to grow is the inverse of this rate; i.e., $t\sim{W\over\sqrt{|B|}}$. In this time, the bands will separate perpendicular to the bands
by a distance 
\begin{align}
|v_v-v_\rho|t\cos\theta\sim{|v_v-v_\rho|\cos\theta W\over\sqrt{|B|}} \,.
\end{align}  
Requiring that this distance be greater than the width $W$ itself leads to the condition:
\beq
{|v_v-v_\rho|W|\cos\theta|\over\sqrt{|B|}}>W
\label{stab cond}
\eeq
or, cancelling off the common factor of the band width $W$,
\beq
{|v_v-v_\rho||\cos(\theta)|\over\sqrt{|B|}}>1 \,.
\label{stab cond 2}
\eeq
This is the condition for stability, because only if this condition is satisfied will the bands separate before the fluctuation grows.
This condition can be rewritten
\beq
|\cos(\theta)|>{\sqrt{|B|}\over|v_v-v_\rho||} \,.
\label{stab cond 3}
\eeq
For small $|B|$ (i.e., near the instability), the right hand side of this inequality becomes very small, so that we can expand the cosine on the left hand side for $\theta$ near $\pm{\pi\over2}$. This leads to the {\it stability} condition
\beq
|\theta\pm\pi/2|>{\sqrt{|B|}\over|v_v-v_\rho|} \,,
\label{stab cond 4}
\eeq
which the alert reader will recognize as equivalent to the {\it instability} condition
(\ref{instab angle}), up to a factor of  $2$, which one wouldn't expect to get correctly from an argument as crude as this one.

\vspace{.2in}

\subsection{Summary of the stability analysis}

In summary, we have shown in this section that autochemotaxis can make the inverse compressibility negative. Next, we demonstrated that {\it any} negative inverse compressibility, no matter how small, always destabilizes the uniform homogeneous flock. However, in contrast to disordered systems, ordered flocks near this instability are only unstable in a narrow region of region of $\bq$ space, very near the direction perpendicular to the direction of flock motion, as illustrated in figure \ref{q-unstable}. 
Furthermore, the wavevector of the {\it most} unstable mode is {precisely} perpendicular to the mean direction of flock motion. 

Finally, we note that the unstable wavevectors are all necessarily small near the threshold of the instability (i.e., when the inverse compressibility  $B$ is negative, but its magnitude $|B|$ is small).

We will use all of the above facts in the next section to help us determine the final steady state the system near, but on the unstable side of, the instability threshold.

\section{The Final state: Phase separation}\label{phase sep}

\subsection{Formulation}\label{form}
We'd now like to determine the final steady state of the system when the instability found in the last section occurs.  We will show in this section that, as claimed in the introduction, this state is ``phase separated": the system separates into two macroscopic bands running parallel to the mean velocity, with essentially constant density within each. The interface between these two bands has a width $\ell$ that is independent of the system width $w$. Therefore, the ratio ${\ell\over w}$ of the interface width to the system width vanishes as system  size $w\to\infty$. The interface width $\ell$ itself diverges as the
 system approaches a non-equilibrium ``critical point" that we will determine.

Since the instability is sharply focused in the directions perpendicular to the bulk motion, we will seek solutions of our equations of motion that depend only on one Cartesian component of position, { by which we mean configurations in which all of the fields depend {\it on only one of the $d-1$ Cartesian coordinates}  perpendicular to $\hx$ ($d$ being the dimension of space).} We'll call this direction $y$. 

In addition, since the instability was in a mode in which only the ``longitudinal" component of the perpendicular velocity $\bvp$ (i.e., the component $v_L$ along the direction of the  projection of the wavevector of the instability  perpendicular to the mean flock velocity vector $\vb$) was non-zero, we will seek solutions in which only that component of $\bvp$ - which, in this geometry, means the y-component- is non-zero. 

Note that in addition there is a fluctuating and position dependent component of the velocity in the $\hat x$ direction. This is due to the density dependence of $U(|\bv|,\rho)$, which, as the reader may recall, maintains the non-zero speed of the flock.

Specifically, we'll search for solutions of the form
\beq
\bvp = v_y(y,t) \hat y \,\,\,\,\,,\,\,\,\,\,
\rho = \rho_{ref} + \Delta\rho(y,t) \,,
\label{1dansatz}
\eeq
where $\rf$ is some reference density. Because we wish to consider the effect of changing the mean density $\rho_0$ of the system, we will chose this reference density to in general differ from  $\rho_0$.
 
 The chemically sensitive reader will notice that we have not included the chemo-attractant concentration $\eta$ in the above analysis. This is because, as we did when we linearized our equations of motion in section IV, we can eliminate this field.  We can similarly eliminate the fluctuating component of the velocity along $\vb$. See section IV for more details.
 
This approach is justified whenever the solutions we are studying vary slowly in space and time. Since we are looking for {\it steady-state} solutions, slow variation in time is guaranteed ({\it no} variation is as slow as you can get!). We will verify {\it a posteriori} that the solutions we find do indeed vary slowly in space, for parameters near a non-equilibrium ``critical point". We will identify this ``critical point" as we proceed.

As discussed in the introduction, our system proves to be closely analogous to 
equilibrium phase separation. In particular, it proves, as the analytic solution we'll present below shows, to have the analog of the liquid-vapor critical 
point\cite{chaikin} of equilibrium phase separation. We will choose our reference density $\rho_{ref}$ to be close to the critical density of our non-equilibrium critical point. This justifies our expansion in powers of $\Delta\rho$, since, as we'll see, it implies that the density $\rho(\br)$  is never very far from $\rho_{ref}$ at any point in space in the phase separated state, provided that the mean density $\rho_0$ is close to $\rho_{ref}$.

We will not go through the process of eliminating the fast variables in detail here. We will content ourselves with noting that
the resultant equations of motion {\it must} be of the form of the Toner-Tu equations (\ref{TTv}) and (\ref{TTrho}). This is an inevitable result of the fact that the symmetries and conservation laws of our problem are the same as those considered in the formulation of the Toner-Tu model.

Note that, in contrast to the stability analysis of the previous section, we will {\it not} linearize our equations of motion. However, we will still truncate them to {\it quadratic} order in the fluctuations $v_y(y)$, to  {\it cubic} order in $\Delta\rho(y)$ , and to {\it bilinear} order in $v_y\Delta\rho$. This will be sufficient if $v_y(y)$ and $\Delta\rho(y)$ are small, as we will again verify they are {\it a posteriori}.

We will further gradient expand our model. In particular, terms involving 2 derivatives of any type ($x$, $\perp$, or $t$) will be expanded only to {\it linear}
order in the fields. This is justified by our expectation that, near the critical  point, the density varies very slowly in space and time. We will verify this assumption {\it a posteriori}.

 The resultant equations read\cite{w2foot}:

\begin{widetext}
\begin{align}
\partial_{t} \bv_\perp + v_v\partial_x 
\bv_\perp + \lambda \left(\bv_\perp \cdot
\nabla_{\perp}\right) \bv_\perp &=-g_1\Delta\rho\partial_x 
\bv_\perp-g_2\bv_\perp\partial_x\Delta\rho  - \nabla_\perp P_1
+ D_{B}\nabla_\perp(\nabla_\perp\cdot\bv_\perp) + D_T\nabla^2_\perp\bv_\perp  \nonumber\\&~+ D_{x}\partial_x^2\bv_\perp+\nu_x\pp_x\nabla_\perp\Delta\rho+\nu_t \pp_t\nabla_\perp\Delta\rho \,,  
\label{vEOMbroken}\\
\partial_t\Delta\rho + \rf\nabla_\perp\cdot\bv_\perp
+\nabla_\perp\cdot(\bv_\perp\Delta\rho)+v_\rho
\partial_x\Delta\rho &= D_{\rho x}\partial^2_x\Delta\rho+D_{\rho\perp}\nabla_\perp^2\Delta \rho+D_{\rho v} \partial_x
\left(\nabla_\perp \cdot \vec{v}_{\perp}\right)\nonumber\\
&~+\phi\partial_t\partial_x\Delta\rho
+w_2\partial_x(\Delta\rho^2)+{\rf\over2v_0}\partial_x(|
\vec{v}_\perp|^2)~, \label{cons broken}
\end{align}
\end{widetext}
where we've defined $\lambda\equiv\lambda_1(\rf, v_0)$.

Inserting our one-dimensional ans\"{a}tze (\ref{1dansatz}) into the Toner-Tu equations (\ref{vEOMbroken}) and (\ref{cons broken}), and dropping terms of higher than quadratic order in $v_y(y)$ and higher than cubic order in $\Delta\rho(y)$ leads to 
PDE's for the one-dimensional evolution:
\begin{align}
   & \partial_t v_y + \lambda v_y\partial_y v_y = D_{_{L\perp}}\partial_y^2v_y - \partial_yP_1 +\nu_t\pp_t\pp_y\Delta\rho
   \label{1.0},\\
    &\partial_t \Delta\rho = D_{_C} \partial_y^2  \Delta\rho 
     - \rf\partial_yv_y-\partial_y(v_y\Delta\rho)  \label{2.0},
\end{align}
where  we've defined
\begin{align}
   D_{L\perp}\equiv D_B+D_T \,\,\,\,\,,\,\,\,\,D_{_C}\equiv D_{\rho\perp} \,.
   \end{align}
In the steady-state, these reduce to two  coupled ODE's:
\beqn
&&\lambda v_y{dv_y\over dy} = D_{_{L\perp}}{d^2v_y\over dy^2} - {dP_1\over dy} \label{1.1},\\
&&D_{_C} {d^2\Delta\rho\over dy^2} 
- \rf{dv_y\over dy}-{d\over dy}(v_y\Delta\rho)=0  \label{2.1} \,.
\eeqn

We will begin by using equations (\ref{1.1}) and (\ref{2.1}) to analytically determine the steady state solution. To validate our solution, we numerically solved the dynamical equations (\ref{1.0}) and (\ref{2.0}), and allowed them to evolve for long times. 

In the previous sections we demonstrated the uniform state with $\rho(\br)=\rho_0$ and $v_y=0$ becomes  unstable when the inverse compressibility $B$ changes sign.  In order to find a stable steady state, it is therefore necessary to extend our expansion of the isotropic pressure $P_1$  to third order in $\Delta\rho$: 

\begin{align}
P_1 = P_0+M\Delta\rho + w \Delta\rho^2 + u \Delta\rho^3 \,.
\label{Pexp}
\end{align}

Recall that, in the linear theory, we found that the inverse compressibility $B$ appearing in the linearized equations of motion become negative for sufficiently strong attractive autochemotaxis. It is therefore clear that the expansion coefficient $M$ in the above expansion will likewise be driven from positive to negative values by increasing autochemotaxis.

We will therefore treat $M$ as an experimentally controllable parameter, which can be decreased by increasing the chemotaxis. The other experimentally controllable parameter at our disposal is the mean density $\rho_0$. 

The experimental situation is thus very similar to studying equilibrium phase separation by tuning both the system's mean density $\rho_0$ and the temperature, with, in our problem, chemotaxis - or its proxy $M$- playing the role of temperature. We will therefore determine the ``phase diagram" of our system in the $\rho_0$-$M$ plane. 

\vspace{.2in}

\subsection{Stability boundary: the ``spinodal" line}

We'll begin by determining the {\it stability boundary} in this $\rho_0$-$M$ plane. As shown in the last section,  the instability occurs when the inverse compressibility vanishes; that is, when
\beq
B\equiv\left({dP_1\over d\rho}\right)\bigg|_{\rho_0}=0 \,.
\label{stab cond P}
\eeq
We will define the value of $M$ that satisfies this condition to be $M_{spinodal}$, since it is the precise analog of the spinodal line in equilibrium phase separation: that is, the line on which the uniform density state becomes unstable.

Inserting our expansion (\ref{Pexp}) into the condition (\ref{stab cond P}) gives 
\beq
\ms
+2w(\rho_0-\rf) + 3u (\rho_0-\rf)^2=0 \,,
\label{spinodal 1}
\eeq
which can easily be solved for the spinodal line:
\beq
\ms=M_c-3u(\rho_0-\rho_c)^2 \,,
\label{ms}
\eeq
where we've defined the ``critical" value $M_c$ of $M$ as
\beq
M_c\equiv{w^2\over3u} \,,
\label{mc}
\eeq 
and the ``critical density"
\beq
\rc\equiv\rf-{w\over3u} \,.
\label{rc}
\eeq
The critical values $M_c$ and $\rc$ of $M$ and $\rho$ prove to be the analogs of the critical temperature and critical density of an equilibrium phase separating system.

Using (\ref{rc}), we can rewrite (\ref{1.1}) and (\ref{2.1}) in terms of the critical density  $\rc$ and a new density variable $\rho'$ defined as the difference between the local density and the {\it critical} density:
\beq
\rho'=\rho-\rc \,.
\label{shift1}
\eeq
Those equations read:
\beqn
&&\lambda v_y{dv_y\over dy} = D_{_{L\perp}}{d^2v_y\over dy^2} - {dP_1\over dy} \label{1.11},\\
&&D_{_C} {d^2{\rho'}\over dy^2} 
- \rc{dv_y\over dy}-{d\over dy}(v_y{\rho'})=0  \label{2.11} \,.
\eeqn

\vspace{.2in}

\subsection{Two-phase coexistence boundary: the ``binodal" line and the common tangent construction}

Now we turn to the question of the phase separated state of the system. In discussing this, it is useful to define the analog $F(\rho)$ of an equilibrium free energy for our system. We can do so by defining
\beq
P_1={dF(\rho)\over d\rho} \,,
\label{F def}
\eeq
which is readily seen to imply
\beq
F(\rho)=F_0+P_0\Delta\rho+{M\over2}\Delta\rho^2 +{w\over3} \Delta\rho^3 + {u\over4} \Delta\rho^4 \,.
\label{fsol}
\eeq

\begin{figure}
        \centering
        \includegraphics[width=\linewidth]{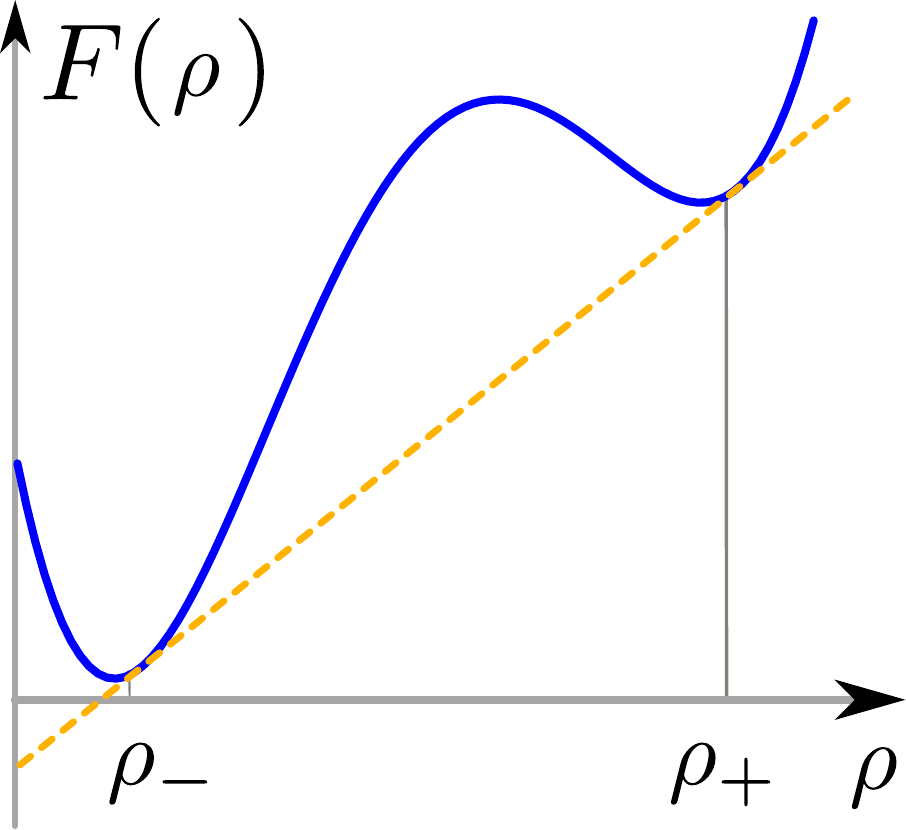}
        \caption{Illustration of the common tangent construction. The densities of the high and low density regions are the points that share a common tangent line. In equilibrium physics, each phase has the same thermodynamic pressure and chemical potential. }
        \label{CTCDemo}
\end{figure}

We will show below that the two densities into which the system ultimately phase separates  in the steady state are determined, {\it sufficiently close to the critical point}
\beq
M=M_c \sep \rho=\rho_c\,,
\label{cp}
\eeq
by the familiar ``common tangent construction" of equilibrium statistical mechanics, applied to the ``free energy" (\ref{fsol}).  This is illustrated in figure \ref{CTCDemo}.

This analogy to equilibrium statistical mechanics {\it only} applies close to the critical point (\ref{cp}). We will also show below that, as we move away from the critical point, the densities are determined by an {\it ``uncommon tangent construction"}: the slope of the free energy (\ref{fsol}) remains the same at both coexisting densities, but the tangents to the free energy curve at those two densities do {\it not} coincide, as illustrated in figure \ref{CTCDemo}.

As pointed out earlier, our calculation here is very similar to Landau theory for equilibrium liquid-gas phase separation\cite{chaikin}. In particular, as in equilibrium Landau theory, our truncation of the expansion of the pressures in powers of $\Delta\rho$ at third order is only justified very near the ``critical point" (\ref{cp}) for our system, and only if the critical density $\rc$ is close to the reference density $\rf$ around which we did our expansion. Of course, since we are free to choose the reference density  $\rf$ to be anything we want, we can {\it always} choose it to be close to the critical density $\rc$. Indeed, we {\it could} have chosen it to be {\it equal} to $\rc$; with that choice, we would have had $w=0$. Hence, if our reference density $\rf$ is close to the critical density, $w$ will be small. We will also restrict our discussion to the vicinity of the critical point (\ref{cp}), which implies that $M$ is small as well.   We will therefore make liberal use of the assumption of small $M$ and $w$ in all of the analysis that follows. 
 
All of these assumptions are precisely the same as those usually made in the Landau theory analysis of an equilibrium critical point for phase separation\cite{chaikin}.  We therefore expect that our analytic results obtained below to be as reliable as those obtained from equilibrium Landau theory: that is, valid (in the absence of fluctuation effects) sufficiently close to the critical point.

For computational convenience, we will shift variable from $\Delta\rho$ to the new density variable $\rho'$ defined in (\ref{shift1}) 
as the difference between the local density and the {\it critical} density. Using our expression (\ref{rc}) for the critical density,  we see that 
\beq
\Delta\rho=\rho'-{w\over3u} \,.
\label{shift}
\eeq
This shift eliminates the quadratic term in the pressure, or, equivalently, the cubic term in the free energy, (\ref{fsol}), enabling us to write
\begin{align}
P_1(\rho')={dF(\rho')\over d\rho'}
\label{Pshift}
\end{align}
with
\beq
F(\rho')= P_0\rho'-{1\over2}m\rho'^{ 2} + {1\over4}u\rho'^{4}+{\rm constant} \,,
\label{fshift}
\eeq
where we've defined
\beq
m\equiv {w^2\over3u}-M \,.
\label{mdef}
\eeq
We have not bothered evaluating the constant in (\ref{fshift}), since it drops out of the equation of motion (\ref{1.11}).

Note also that since we are limiting our analysis to small $M$ and $w$, $m$ is also small. We will make use of this fact in the perturbation theory we develop for the non-equilibrium phase separation later.

With the expansion (\ref{fshift}) in hand, we now further simplify  (\ref{1.11}) and (\ref{2.11}) by rescaling to dimensionless coordinates and variables, as follows:
\beq
s\equiv {y\over \ell}  \,\,\,\,\,\,,\,\,\,\,\, \tilde{v}\equiv {v_y\over v_s}  \,\,\,\,\,\,,\,\,\,\,\, \tilde{\rho}\equiv {\rho'\over \rho_s} \,.
\label{dless}
\eeq 
with the characteristic length $\ell$, speed $v_s$, and density $\rho_s$ given by 

\begin{align}
    \ell &= \sqrt{\frac{D_{_{L\perp}}D_{_C}}{\rc m}} \,,\label{elldless}\\
    v_s &= \frac{m\sqrt{D_{_C}}}{\sqrt{D_{_{L\perp}} \rc u}} \,,\\
    \rho_s &= \sqrt{\frac{m}{u}} \,.     \label{dlesscoefficients}
\end{align}
We'll see shortly that the length $\ell$ is roughly the width of the interface between phase separated regions, while the difference between the densities of the two phase separated regions is roughly $\rho_s$.

In terms of these new coordinates and variables, our dimensionless steady-state equations of motion (\ref{1.11}) and (\ref{2.11}) read:

\begin{align}
   {d^2 \tilde{v}\over ds^2}  &= {d\over ds}\left({d\tf(\tr)\over d\tr}\right) + \Lambda \tilde{v} {d \tilde{v}\over ds}  \label{nondimFull1}\,,\\ 
   {d^2 \tilde{\rho}\over ds^2} &= {d \tilde{v}\over ds} + \Gamma {d\over ds}(\tilde{\rho} \tilde{v}) \,,\label{nondimFull2}
\end{align}
where we've defined the dimensionless parameters: 
\begin{align}
    \Lambda &= \frac{\lambda D_c}{D_{_{L\perp}}\rc}\sqrt{\frac{m}{u}} =\frac{\lambda  D_c\rho_s}{D_{_{L\perp}}\rc}\,,\\
    \Gamma &= \frac{1}{\rc}\sqrt{\frac{m}{u}} ={\rho_s\over\rc}\,,
\end{align}
and the dimensionless free energy
 \beq
 \tf(\tr)\equiv{\ell\over\rho_s v_sD_{_{L\perp}}}F(\rho'=\rho_s\tr)=\tp\tr-{1\over2}\tr^2+{1\over4}\tr^4 \,.
 \label{ftildedef}
 \eeq
   In (\ref{ftildedef}), we've defined the dimensionless pressure 
 \beq
 \tp\equiv{P_0\ell\over D_{_{L\perp}}v_s} \,.
 \label{dlesspdef}
 \eeq
In doing this rescaling, we have implicitly assumed that $m$ is positive ($m>0$), as it obviously will be for $M<M_c$ (see equation (\ref{mdef}).

Note that, because $m$ is small near the critical point, both of our dimensionless parameters $\Gamma$ and $ \Lambda$
are also small near the critical point. Therefore, to leading order near the critical point, we can set them to zero. Doing so gives, for the steady state,
\begin{align}
    {d^2 \tilde{v} \over ds^2} &={d\over ds}\left({d\tf(\tr)\over d\tr}\right) \,, \label{nondimFull10}
    \\ 
    {d^2\tilde{\rho}\over ds^2} &= {d\tilde{v}\over ds}  \,.\label{nondimFull20}
\end{align}
Differentiating (\ref{nondimFull20}) with respect to $s$ and using equation (\ref{nondimFull10}) in the result gives
\begin{align}
    {d^3 \tilde{\rho}\over ds^3} &= {d\over ds}\left({d\tf(\tr)\over d\tr}\right) \label{linbandsEquation}.
\end{align}
Integrating this once with respect to $s$ then gives
\beq
{d^2\tilde{\rho}\over ds^2}  = \left({d\tf(\tr)\over d\tr}\right)  + \mu \,,
\label{linbandsEquation2}
\eeq 
where $\mu$ is a constant of integration. 
Multiplying both sides of (\ref{linbandsEquation2}) by  $\frac{d\tilde\rho}{ds}$
and integrating with respect to $s$ again reveals that this equation has a conserved  first integral of motion. That is, the quantity
\beq
E\equiv -{1\over2}\left({d\tilde{\rho}\over ds}\right)^2+\tilde{F}(\tilde{\rho})+\mu\tilde{\rho}
\label{Econs}
\eeq
is a constant (independent of position $s$).

Assuming that the density profile in the steady state consists of a set of ``plateaus" - i.e., regions of essentially constant density - we can calculate the possible plateau densities, which we'll call $\tilde{\rho}_\pm$ (with $\tr_-$ being the smaller of the two), from equations (\ref{linbandsEquation2}) and (\ref{Econs}). The result, as we'll now show, is identical to the familiar ``common tangent construction" of equilibrium phase separation.

On a plateau, all spatial derivatives must vanish. Hence, 
(\ref{linbandsEquation2}) implies
\beq
{d\tf\over d\tilde\rho}\bigg|_{\tilde\rho_{\pm}}=-\mu \,.
\label{common slope}
\eeq
That is, at the plateau densities, the derivatives of the effective free energy $\tf$ must be equal. Since $\tf$ and $\tr$ are linearly related to to our dimensionful free energy $F(\rho)$ and $\rho$, this condition clearly also applies to the derivatives of $F$ with respect to $\rho$.

Now setting derivatives equal to zero in (\ref{Econs}), and using the fact that the ``energy" $E$ must be the same on both plateaus, we obtain  a relation between the values of $\tf$ itself (as opposed to its derivatives) on the plateaus:
\beq
\tf(\tilde{\rho}_+)+\mu\tilde{\rho}_+=\tf(\tilde{\rho}_-)+\mu\tilde{\rho}_- \,.
\label{common tangent 1}
\eeq
This expression can be reorganized to read
\beq
\tf(\tilde{\rho}_+)=\tf(\tilde{\rho}_-)-\mu(\tilde{\rho}_+-\tilde{\rho}_- )\,.
\label{common tangent 2}
\eeq
Using the fact that (\ref{common slope}) implies that ${d\tf\over d\tilde\rho}|_{_{\tilde\rho_{\pm}}}=-\mu$, we see immediately that (\ref{common slope}) and (\ref{common tangent 2}) taken together imply the ``common tangent construction": a tangent to a plot of $\tf(\tilde{\rho})$ at one of the plateau densities $\tilde{\rho}_\pm$ must also be tangent to the plot at the other plateau density. 
As for the common slope relation, here too,  this relation must apply to the original, unrescaled density $\rho$ and free energy $F$ as well.

This common tangent construction is illustrated in figure \ref{CTCDemo}.

To solve for these two steady state densities, it is most convenient to define a new ``pseudo-Gibbs free energy" $G$ via
\beq
G(\tr)\equiv \tilde{F}+\mu\tr=(\tp+\mu)\tr-{1\over2}\tr^2+{1\over4}\tr^4 \,.
\label{Gdef}
\eeq
Keep in mind that $\mu$ is an as yet undetermined constant of integration, which we must adjust in order to obtain plateaus in the density.

With the definition (\ref{Gdef}), we see that the condition (\ref{common slope}) on the slope of the effective free energy $\tf$ at the plateau densities becomes the simpler condition
\beq
{dG\over d\tr}\bigg|_{\tilde\rho_{\pm}}=0 \,,
\label{Gcond}
\eeq
i.e., $G$ must be minimized (or maximized) at $\tr_\pm$.

Furthermore, the condition (\ref{common tangent 1}) implies
\beq
G(\tilde{\rho}_+)=G(\tilde{\rho}_-) \,.
\label{common tangent G}
\eeq
Taking the conditions (\ref{Gcond}) and (\ref{common tangent G}) together implies that the arbitrary constant of integration $\mu$ must be chosen so that $G$ has two {\it degenerate} minima. Inspection of (\ref{Gdef}) makes it fairly obvious what choice of $\mu$ will accomplish this:
\beq
\mu=-\tp \,.
\label{musol}
\eeq

With this choice, we have
\beq
G(\tr)=-{1\over2}\tr^2+{1\over4}\tr^4 \,,
\label{Gnomu}
\eeq
which clearly has two degenerate minima at

\beq
\tilde{\rho}_\pm=\pm 1 \,.
\label{rhopm}
\eeq
Undoing the rescaling (\ref{dless}) and the shift (\ref{shift1}), we see that this implies the two plateau densities $\rho_\pm$ are given by
\beq
\rho_\pm=\rc\pm\rho_s=\rc\pm\sqrt{\frac{m}{u}} \,.
\label{platden}
\eeq
Note that, for small $m$ (i.e., close to the critical point), these densities are close to the critical density $\rc$, which justifies our expansion in $\Delta\rho$.

However, the two phase separated states only become accessible if the mean density $\rho_0$ lies between these two phase separated densities:
\beq
\rho_+>\rho_0>\rho_- \,\,\,\,\,.
\label{bincond1}
\eeq
Otherwise, conservation of particle number forbids going to a phase separated state (since both plateaus would have either have a {\it lower} density than the mean density $\rho_0$, or a higher density). In either case, there is no way that one can achieve the mean density $\rho_0$ by putting part of the system in a plateau at $\rho_+$, and the remainder in a plateau of density $\rho_-$. It is only in the regime (\ref{bincond1}) that such a separation can lead to the mean density $\rho_0$.

Using our result (\ref{platden}) for $\rho_\pm$, we see that the condition (\ref{bincond1}) is equivalent to
\beq
\sqrt{\frac{m}{u}}>|\rho-\rc| \,.
\label{bincond2}
\eeq
Using our expression (\ref{mdef}) relating $m$ and the parameter $M$ in our expansion (\ref{Pexp}) for the isotropic pressure $P_1$, we see that this condition can be written as
\beq
M<{w^2\over3u}-u(\rho-\rc)^2\equiv M_{\rm binodal} \,.
\label{bincondf}
\eeq
Note that $\mb>\ms$ for all $\rho$. In fact, the ratio
\beq
{(\ms-M_c)\over(\mb-M_c)}=3
\label{Mrat}
\eeq
is universal for all $\rho$ sufficiently close to the critical point.

\begin{figure}
        \centering
        \includegraphics[width=\linewidth]{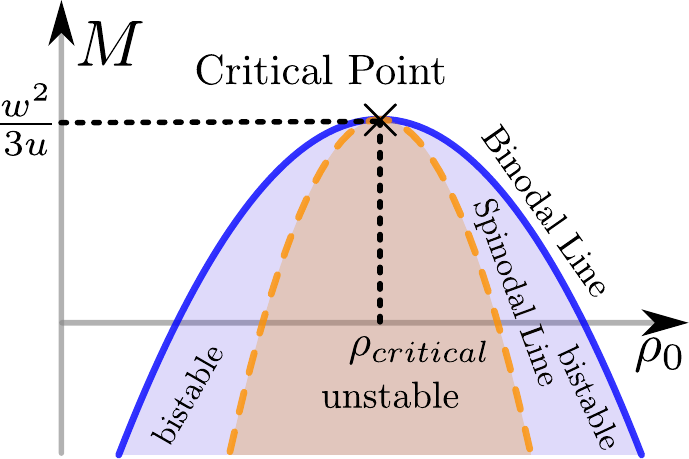}
        \caption{Phase diagram of flock phase separation. The solid blue and  dashed orange curves are the binodal and spinodal lines respectively , analytic expressions for which are given by equations (\ref{bincondf}) and (\ref{ms}). Note that those expressions are only valid close to the critical point. In the orange filled region  under the spinodal line, which is labeled ``unstable", only the two-phase state is stable. In the blue region between the spinodal and binodal lines, both the two phase state, and the homogeneous, one phase state, are stable. The analysis we perform in this paper is done close to the critical point, where our assumption that density variations are small is valid.}
        \label{spinodal}
\end{figure}

The result (\ref{bincondf}), along with our earlier result (\ref{ms}) for the ``spinodal" value $M_{\rm spinodal}$ of $M$ at which the uniform state becomes unstable, can be summarized by the ``phase diagram" in the  $\rho_0$-$M$ plane illustrated in figure \ref{spinodal}. For all $M$ above the ``binodal" parabola, which is the upper parabola in \ref{spinodal}, the system can only be in the uniform state, which is stable. For values of $M$ between the two parabolae (the blue and orange  curves in figure \ref{spinodal}) - that is, for $\mb>M>\ms$, with $\ms$ and $\mb$ given by (\ref{ms}) and (\ref{bincondf}),  
both the two phase state, and the homogeneous, one phase state, are stable.
Finally, for $M<\ms$, only the phase separated state is stable.

The strong similarity between our results and equilibrium liquid-vapor phase separation is apparent from this phase diagram, which is identical to that for an equilibrium liquid-vapor system, with $M$ playing the role of temperature. Recall that $M$ can be increased (decreased) experimentally by decreasing (increasing) the strength of the chemotaxis, or, more generally, by tuning the strength of whatever attractive interactions in the flock reduce the inverse compressiblity. 

Note there is an important difference between our results and 
equilibrium liquid-vapor phase separation. In {\it equilibrium} systems, a homogeneous density state is only  {\it meta}-stable in the region between the binodal and spinodal curves. The homogeneous state exists at a {\it local} minimum in the free energy; the {\it global} minimum is the phase separated state. In the presence of sufficiently large fluctuations, the homogeneous phase will overcome this energy barrier  through droplet nucleation\cite{chaikin}, which leads to phase separation. Since flocks are {\it non-equilibrium} systems, we do not have any reason to believe that the global minimum of our pseudo-energy ($G$) corresponds to a ``preferred" stable state. All we can determine is the {\it local} stability of each phase between the binodal and spinodal curves. Between these two curves, both states 
are locally stable. We therefore refer to this region as the bistable region.

\begin{figure}
        \centering
        \includegraphics[width=\linewidth]{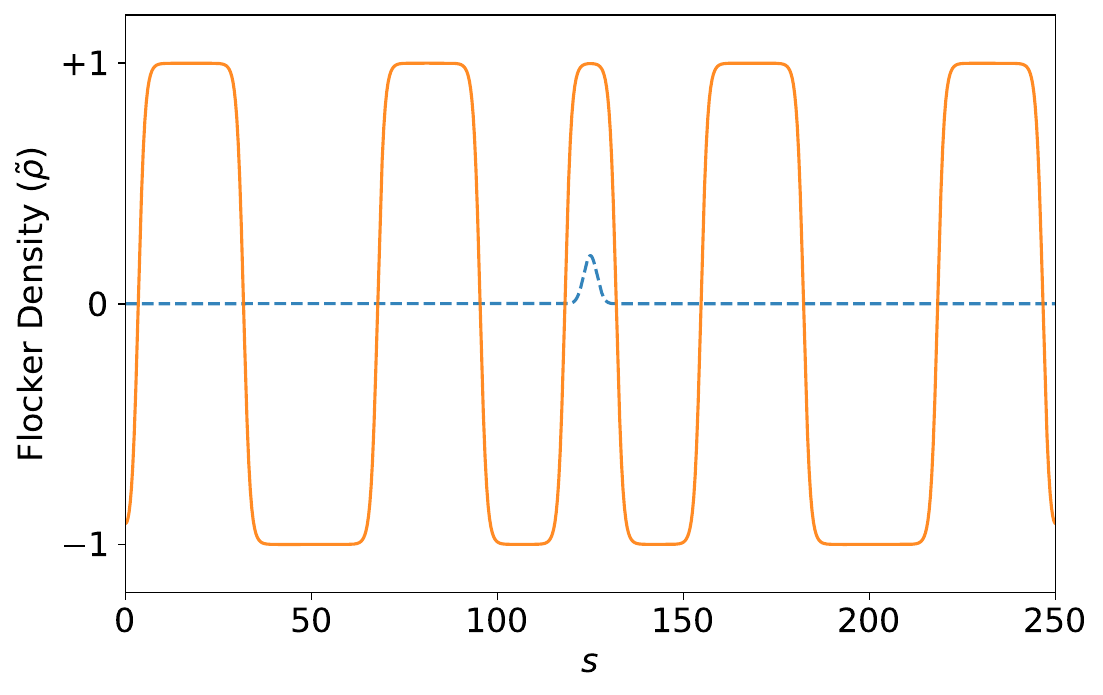}
        \caption{A result from evolving equation (\ref{nondimFull1}) and (\ref{nondimFull2})
        for a long time, in a periodic space. The dashed blue curve is the initial condition, and the orange curve is the final state. This solution was iterated for 3000 units of time and $\Lambda, \Gamma =0$. The plateaus are very weakly attracted to one another.  The long time steady state solution in a periodic space would consist one a single high density region and a single low density region. }
        \label{LinearBands}
\end{figure}

We turn now from the plateau densities to the form of the interface between different plateaus. We will determine this using equation (\ref{Econs}). Using our earlier result that $\mu=-\tp $, and evaluating the right hand side of (\ref{Econs}) on a plateau, where $\tilde{\rho}=\pm1$ and ${d\tilde{\rho}\over ds}=0$, we see that 
\beq
E=-{1\over4} \,.
\label{E}
\eeq
Using this in  (\ref{Econs}), we can solve that equation for ${d\tilde{\rho}\over ds}$:
\beq
{d\tilde{\rho}\over ds}=\pm \sqrt{{1\over2}(\tilde{\rho}^4-2\tilde{\rho}^2+1)}
=\pm{1\over\sqrt{2}}(1-\tilde{\rho}^2) \,,
\label{dsrho}
\eeq
where the plus sign corresponds to the transition region from the low density plateau at $\tilde{\rho}=-1$ on the left to the high density plateau at $\tilde{\rho}=1$ on the right, and the minus sign to the opposite case. Solving this first order ODE by separation of variables, we find the interface structure:
\beq
    \tilde{\rho}(s)=\pm\tanh\left({s-s_0\over\sqrt{2}}\right) \,,
 \label{wall}
 \eeq
where the constant of integration $s_0$ simply gives the position of the interface.

Undoing our various shifts (\ref{shift})  and rescalings (\ref{dless}), we can use this to write the   density profile of the interface in terms of the physical variable $\rho$ and the physical coordinate $y$:
\beqn
\rho(y)
&=& \rf-{w\over3u}\pm\sqrt{\frac{{w^2\over3u}-B}{u}}\tanh\left({y-y_0\over\ell\sqrt{2}}\right) \,,
\nonumber\\
\label{wallreal}
\eeqn
where as before the plus sign describes the case in which the low density plateau is on the left (i.e., as $y\to-\infty$) and the high density on the right, while the minus sign describes the opposite case.

Note that, as claimed earlier, the characteristic width of these interfaces is of order $\ell$, and hence diverges like $m^{-1/2}$ as we approach the critical point $m=0$, as can be seen from equation (\ref{elldless}).

Having found the density $\rho$, we can now also determine the velocity field $\bv$. The $y$-component follows from equation (\ref{nondimFull20}), which can be integrated once to give
\beq
\tilde{v}={d\tilde{\rho}\over ds}+{\rm constant}
=\pm{1\over\sqrt{2}\cosh^2\left({s-s_0\over\sqrt{2}}\right)}+{\rm constant} \,.
\label{vy1}
\eeq
The constant of integration in this expression is readily seen to be zero, since, as can be seen from our full time-dependent 1d equations of motion (\ref{1.1}) and (\ref{2.1}) the volume integral of $\tilde{v}$ is conserved. This volume integral was zero initially, because we started with a system moving uniformly in the $x$-direction, so $v_y{(y, t=0)}=0$ for all $y$. Hence, this volume integral must be zero in the final state, as it will be if we choose the constant of integration in (\ref{vy1}) to be zero (note that a system with periodic boundary conditions must have equal numbers of ``up" and ``down" interfaces, so the small contributions to $\int dy v_y(y)$ coming from these regions will cancel).

\begin{figure}
    \centering
    \includegraphics[width=\linewidth]{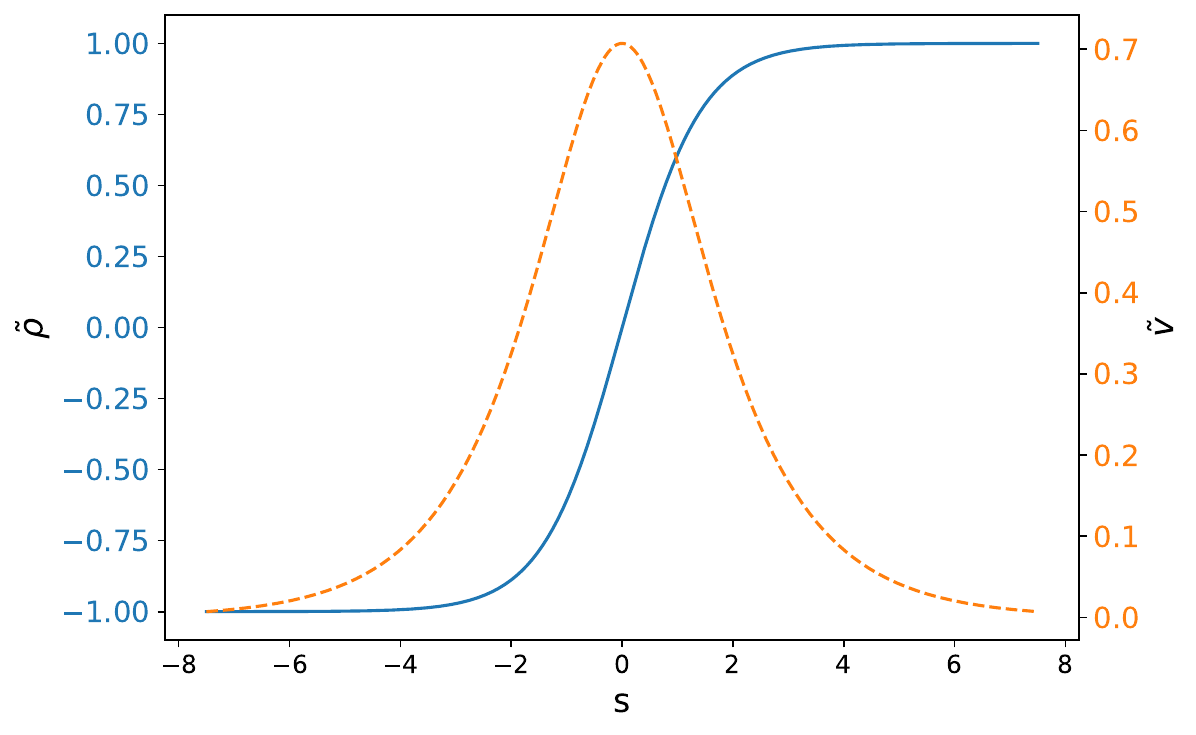}
    \caption{
   Interface density and velocity profile, as described by plus variants of (\ref{wall}) and (\ref{vy2}), where $s_0=0$. The blue line is the density fluctuation. The dashed orange line is the velocity field. Boids accelerate from a low density region (on the left) and move rightward to the high density region. They reach their maximum velocity halfway through the interface and begin to decelerate as they move deeper into the plateau. This accumulation of boids is balanced with diffusion from the high density region to the low density region.}
    
    \label{DensityVelocityPlot}
\end{figure}

Thus we have 
\beq
\tilde{v}
=\pm{1\over\sqrt{2}\cosh^2\left({s-s_0\over\sqrt{2}}\right)} \,.
\label{vy2}
\eeq
Undoing our rescaling (\ref{dless}), we see that the true $y$-component of the velocity is given by
\beq
v_y(y)
=\pm{v_s\over\sqrt{2}\cosh^2\left({y-y_0\over\ell\sqrt{2}}\right)} \,,
\label{vyreal}
\eeq
which shows that the velocity is localized to within a distance $\ell$ of the interface, and that its characteristic velocity scale $v_s$ vanishes like $m$ as the critical point is approached.
The density and the velocity profiles (\ref{wall}) and (\ref{vy2}) are plotted in figure \ref{DensityVelocityPlot}. 

The steady state profile of the interface is stabilized by a balance between the velocity field (\ref{vyreal}), which convectively carries boids from the low density region to the high density region, and diffusion (controlled by $D_{\rho\perp}$) which carries boids from the high density region to the low density region.

Finally, we note that the $x$-component of the velocity must also become position dependent; in particular, it must take on different values in the plateaus. This is true simply because, in general, the mean local speed of the flock should depend on the local density. Since that density is different in the two plateaus, the speeds in each plateau should be different as well. 

This can be verified by looking at the $x$-component of the Toner-Tu-Keller-Segel equations (\ref{KSTTvIntro}-\ref{KSTTetaIntro}) in each plateau. In a plateau, all spatial derivatives must vanish. In addition, in the steady state, time derivatives must vanish as well. Hence, the $x$-component of the Toner-Tu-Keller-Segel equations reduces to
\beq
U(|\bv|,\rho, \eta) =0 \,.
\label{U0}
\eeq
This is an implicit equation relating the speed $|\bv|$ to the density $\rho$ and 
chemo-attractant concentration $\eta$. Since $v_y=0$ in the plateaus (as we just saw), $|\bv|=v_x$ there. Hence, $v_x$ will be a function of $\rho$ and  $\eta$. And since $\rho$ and  $\eta$ take on different values in the plus and minus plateaus, the speed $v_x$ will take on different values as well.

For example, in models like the Vicsek model, in which the speed is an increasing function of $\rho$, the birds in the high density plateaus will be moving (parallel to the plateau boundaries) {\it faster} than those in the low density plateaus. For systems exhibiting jamming-like behavior, in which the local speed is a {\it decreasing} function of the local density, the opposite will be true.

All of the above predictions for the final steady state of the system are fully borne out by 
numerical solution of the time-dependent equations (\ref{nondimFull1}) and (\ref{nondimFull2}) in a periodic space,  as illustrated in figure \ref{LinearBands}, and in the animation Movie 1, in the supplemental materials. 

All of our numerical solutions  started with initial conditions in which the departure from a uniform state was small.  Furthermore, we performed the analog of a ``critical quench" in an equilibrium system, by choosing our mean density to be the critical density $\rc$.

These simulations all show that at small times, the system quickly forms multiple well-defined plateaus, with density and velocity profiles that agree well with our analytic solution. Adjacent plateaus merge quickly, while distant plateaus do not merge on time scales numerically accessible to our  numerical solution (and patience). 

This slowing down is a consequence of the exponential approach to the uniform plateau 
state of both the density (as shown by equation (\ref{wall})) and the velocity (as shown by equation (\ref{vy2})). It is only the overlap of these exponential tails that leads to any interaction at all between neighboring interfaces. As a result, the attraction between neighboring interfaces falls off exponentially with their separation. Hence, the time for neighboring domains of high density to merge also grows exponentially with their separation, which is why the merger slows down so dramatically once the near-lying plateaus have merged.

We believe that this slowing down is an artifact of both the one-dimensionality of our numerical solutions, and the fact that they are noiseless. As discussed in the introduction, we expect that what will really happen, once the plateaus form, is that the bands will begin to undulate due to noise in the system, as illustrated in figure \ref{noiseBands}. These undulations will grow with time until bands begin bumping into  their neighbors, at which point the bands can start to merge. The merged region will then rapidly ``zipper" along in both directions, merging the two bands.  This process will then repeat with the larger bands formed as a result of such mergers, until full phase separation is achieved.

Again reiterating our discussion in the introduction, we expect  the size of these undulations to grow algebraically with time, as the interface fluctuations in the KPZ equation do\cite{KPZ1}. Therefore, the time scale needed for undulations of the bands to grow large enough to reach neighboring bands will grow only {\it algebraically} with the separation between bands, rather than exponentially. We therefore expect that the time required for our autochemotaxic instability to completely phase separate a large system will grow only algebraically with system size.

To make these ideas precise, a theory of the dynamics of fluctuations of an interface between two flocks of different densities and speeds. This would be an excellent topic for further research.

Indeed, a quantitative comparison of our analytical result, (\ref{wall}) and (\ref{wallreal}), for the shape of the interfaces and plateau densities fits the final steady state of these numerical solutions extremely well. 

\vspace{.2in}

\subsection{Two-phase coexistence boundary further from the critical point: the  {\it Uncommon} Tangent Construction}

The previous section's analysis neglected all non-linearities in the equations of motion except for those in the isotropic pressure $P_1$. As we saw, this reduces the statics of the problem to those of the equilibrium system. 

While, as we also saw in the previous section, it is asymptotically exact to neglect those other non-linearities as we approach the critical point, it must also be the case that all non-equilibrium effects, at least on the plateau densities and the binodal and spinodal curves, must come from those neglected non-linearities. To see those effects, we must therefore go beyond the leading order behavior near the critical point, and include the extra non-linearities 
$\Lambda$ and $\Gamma$ in equations (\ref{nondimFull1}) and (\ref{nondimFull2}).  

Recall that we are always limited in any event to analyzing the system close to the critical point, just as one is in equilibrium, because only near the critical point are the density variations small enough to justify our expansion (\ref{Pexp}) of the isotropic pressure. Therefore, we will continue to assume that we're close to the critical point, where $\Lambda$ and $\Gamma$ are small. This justifies treating $\Lambda$ and $\Gamma$ as small perturbations to the problem, and calculating their effects perturbatively. 

We will now do such a perturbation theory, and show that the effect of those terms is to change the {\it common} tangent construction of the previous section to an {\it uncommon} tangent construction: the condition that the {\it slopes}  of our effective free energy at the two plateau densities be equal survives, but the tangent at either of these no longer intersects the free energy curve at the other.

A very similar ``uncommon" tangent construction was found\cite{mipsreview} for ``motility induced phase separation",  which occurs in some {\it disordered} active systems.

 In the steady state, our full equations (\ref{nondimFull1}) and (\ref{nondimFull2}) can be rewritten:
\begin{align}
   {d^2 \tilde{v}\over ds}  &= {d\over ds}\bigg(\partial_{\tilde{\rho}} \tf(\tilde{\rho}) \bigg)+ {\Lambda\over2} 
   {d \tilde{v}^2\over ds}  \label{nondimFull3}\,,\\ 
   {d^2 \tilde{\rho}\over ds^2} &= {d \tilde{v}\over ds} + \Gamma {d\over ds}(\tilde{\rho} \tilde{v}) \,,\label{nondimFull4}
\end{align}
where $\tf(\tilde{\rho})$ is the effective free energy we introduced in equation (\ref{ftildedef}).

Just as in the common tangent construction, we take the derivative of the second equation (\ref{nondimFull4}), and substitute the first equation  (\ref{nondimFull4})  into it. This gives
\begin{align}
    {d^3 \tilde{\rho}\over ds^3} &={d\over ds}\bigg(\partial_{\tilde{\rho}} \tf + \frac{\Lambda}{2}\tilde{v}^2 \bigg)+ \Gamma {d^2\over ds^2}(\tilde{\rho}\tilde{v}) \,.
\end{align}
Next, we rearrange this equation, keeping the new nonlinearities on the right hand side. Each term is a derivative with respect to $s$, so we'll integrate once, obtaining
\begin{align}
    \bigg[{d^2\tilde{\rho}\over ds^2} -\partial_{\tilde{\rho}} \tf(\tilde{\rho})\bigg] ={ \frac{\Lambda}{2}\tilde{v}^2 }+ {d\over ds} \bigg(\Gamma (\tilde{\rho} \tilde{v})\bigg) + \mu \,,\label{useFordeltalaterequation}
\end{align}
where $\mu$ is a constant of integration. Next, we multiply each side by ${d\tilde{\rho}\over ds}$.  This enables us to rewrite (\ref{useFordeltalaterequation}) in the form
\bew
\begin{align}
    {d\over ds}\bigg[\frac{1}{2}\bigg({d\tilde{\rho}\over ds}\bigg)^2 -G(\tilde{\rho})\bigg] ={\frac{\Lambda}{2}\bigg({d\tilde{\rho}\over ds}\bigg)\tilde{v}^2} + \Gamma \bigg({d\tilde{\rho}\over ds}\bigg){d\over ds}\bigg(\tilde{\rho} \tilde{v}\bigg) \,,
    \label{near cons}
\end{align}
\ew
where $G$ is our ``pseudo-Gibbs free energy", given by equation (\ref{Gdef}); that is $G(\tilde{\rho}) = \tf(\tilde{\rho}) + \mu\tilde{\rho}$. 

\begin{figure}
        \centering
      \includegraphics[width=\linewidth]{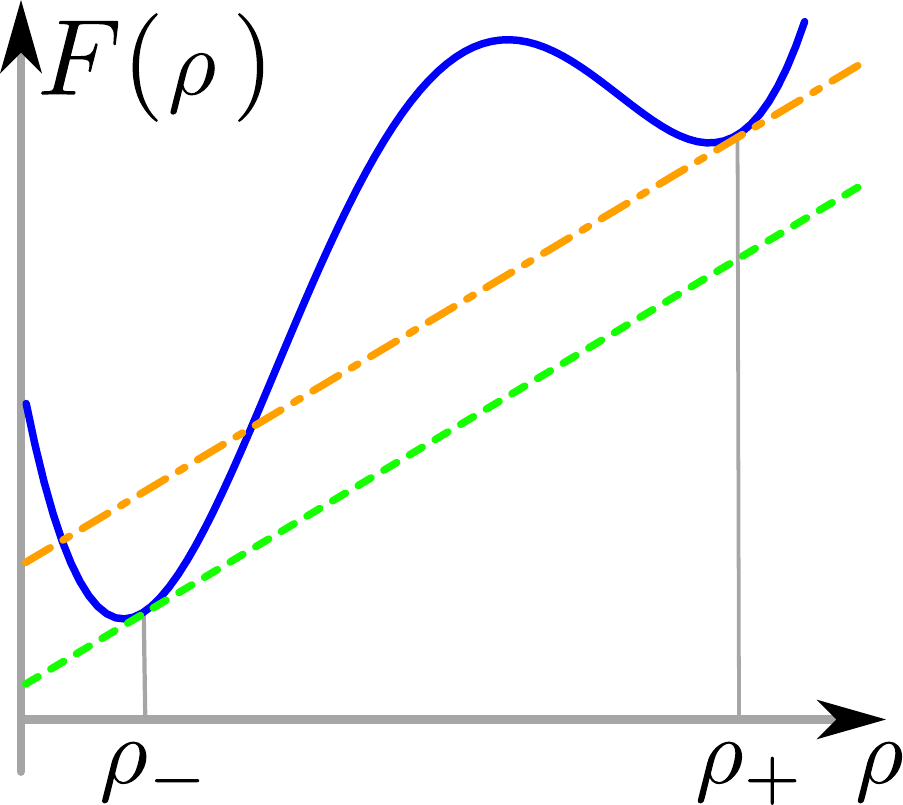}
        \caption{Illustration of the uncommon tangent construction. The {\it slope} of the pseudo-Free energy $F(\rho)$ is the same at both plateau densities $\rho_\pm$, but the two tangents are distinct, rather than being the same line, as they are in equilibrium.  It is only in the limit as the system approaches the critical point that the common tangent construction is recovered.}
        \label{UTCDemo}
\end{figure}

Note that if we set the non-linear coefficients $\Lambda$ and $\Gamma$ to zero, this recovers our earlier result that the term in square brackets on the left hand side of (\ref{near cons}) is a constant. That constancy then leads directly to the common tangent construction of the last section.

Now we wish to work to next order in the  non-linear coefficients $\Lambda$ and $\Gamma$.
We'll do this by writing
\begin{align}
    \tilde{\rho} &= \tilde{\rho}_1 + \tilde{\rho}_2,\\
    \tilde{v} &= \tilde{v}_1 + \tilde{v}_2.
\end{align}
where $\tilde{\rho}_1$ and $\tilde{v}_1$ are the common tangent solutions (\ref{wall}) and (\ref{vy2}) found in the previous section, and $\tilde{\rho}_2$ and $\tilde{v}_2$ are small perturbations to those solutions;  that is

\beq
\tilde{\rho}_1= \pm\tanh\left({s-s_0\over\sqrt{2}}\right),\,
\label{tr1def.1}
\eeq
\beq
\tilde{v}_1= \pm{1\over\sqrt{2}\cosh^2\left({s-s_0\over\sqrt{2}}\right)}. \,
\label{vt1def.2}
\eeq

We'll assume, and verify {\it a posteriori}, that 
$\tilde{\rho}_2$ and $\tilde{v}_2$ are of order $\Lambda$ and $\Gamma$; $\tilde{\rho}_1$ and $\tilde{v}_1$ are, obviously, zeroeth order 
in $\Lambda$ and $\Gamma$.

In light of this perturbative hierarchy, we can, to next order in  $\Lambda$ and $\Gamma$, replace $\tilde{\rho}$ and $\tilde{v}$ with $\tilde{\rho}_1$ and $\tilde{v}_1$ on the right hand side of 
(\ref{near cons}). Doing so, using $\tilde{v}_1 = \partial_s\tilde{\rho}_1$, and integrating over $s$ from a low density plateau to a high density one, and remembering once again that within a plateau, ${d\tr\over ds}=0$, gives
\bew
\beqn
    G(\tilde{\rho}_+)-G(\tilde{\rho}_-) &=& -\frac{\Lambda}{2}{ \int_{-\infty}^\infty}
   \bigg({d\tilde{\rho}_1\over ds}\bigg)^{ 3}ds  -\Gamma{ \int_{-\infty}^\infty}  \bigg({d\tilde{\rho}_1\over ds}\bigg){d\over ds}\bigg(\tilde{\rho}_1 \left({d\tilde{\rho}_1\over ds}\right)\bigg)ds \,.
    \label{Gdif}
\eeqn
\ew
Plugging (\ref{tr1def.1}) into this expression, and  performing the resulting elementary integrals, we find: 
\begin{align}
    G(\tilde{\rho}_+)-G(\tilde{\rho}_-) &= 2\tilde{P}_0+ 2\mu = -\frac{4}{15}[\Lambda+\Gamma]\\
    \mu &= -\tilde{P}_0-\frac{2}{15}[\Lambda+\Gamma].
    \label{FPT}
\end{align}

Since we are working in the  limit in which the nonlinear coefficients are all small, we see that (\ref{FPT}) implies that $\mu+\tilde{P}_0$ is small. We will make use of this fact in a moment.

 Next, we evaluate equation (\ref{useFordeltalaterequation}) deep in the density plateaus, where all spatial derivatives vanish. This implies:
\begin{align}
   -\frac{\partial \tilde{F}(\tilde{\rho}_\pm)}{\partial\tilde{\rho}} =-\tilde{P}_0+\rho_\pm-\rho_\pm^3= \mu \,.
\end{align}

We define $\delta_\pm = \tilde{\rho}_\pm \mp 1$ as the small deviation of these densities from those we obtained earlier by ignoring these ``new" non-linearities. From equation (\ref{FPT}), we see that these obey
\begin{align}
    -(\pm 1 + \delta_\pm) + (\pm1 + \delta_\pm)^3 &= -\mu -\tilde{P}_0\,,
    \label{deltacond}
    \end{align}
    which is easily solved  to linear order in the small quantities $\delta_\pm$ to give
    \begin{align}
    \delta_\pm = -\left(\frac{\mu+\tilde{P}_0}{2}\right) = \frac{1}{15}[\Lambda+\Gamma] \,.
    \label{delta}
\end{align}

Counter-intuitively, the shift is the same for the low and high density plateaus. We strongly suspect that this does not hold to higher orders in perturbation theory, although we have not attempted such a calculation. 

\begin{figure}
        \centering
        \includegraphics[width=0.8\linewidth]{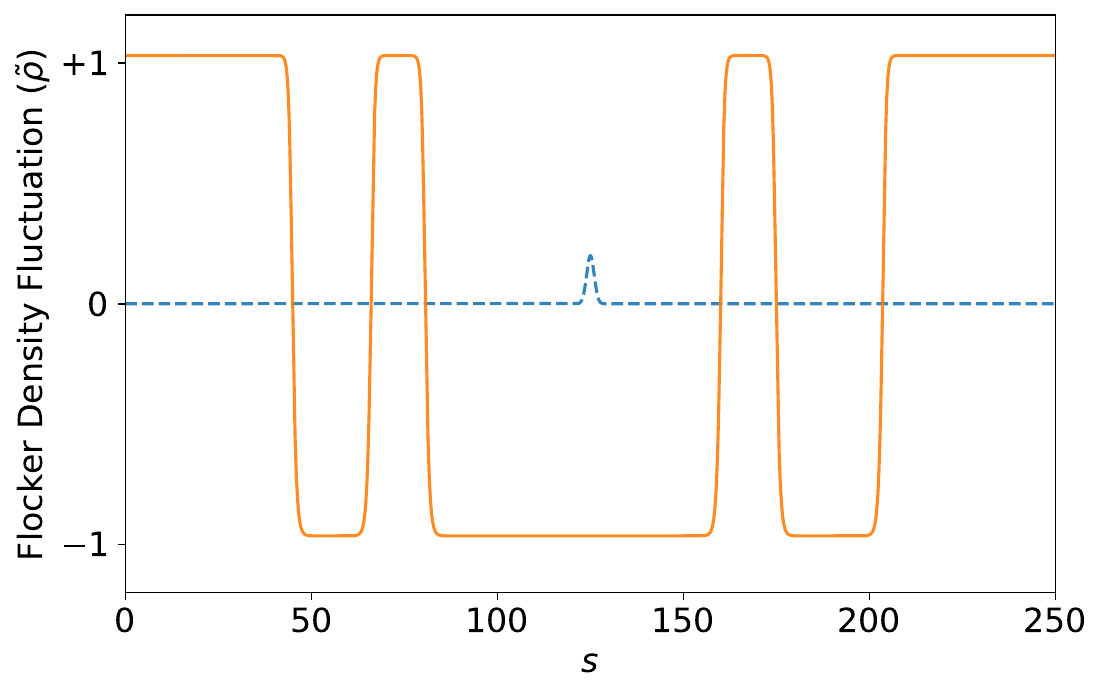}
        \caption{The result of evolving equations  (\ref{nondimFull3}) and (\ref{nondimFull4}) for a long time in a periodic space. The dashed blue curve  is the initial state of the numerical solution. The orange curve is the density profile once the final time step is reached. We retain the plateau structure but it is shifted upwards. The predicted offset $\delta_\pm$  differs from the result obtained by direct numerical solution of the equations of motion, (\ref{1.11}) and (\ref{2.11}), by less than 6\%. The parameters used for this numerical solution were $\Gamma,\Lambda=0.25$ and it was iterated for 3000 units of time.}
        \label{NonLinearBands}
\end{figure}

Figure \ref{NonLinearBands} shows the excellent  agreement between the  analytically predicted offset (\ref{delta}) and the results of simulating the time dependent equation of motion for a long time.

An uncommon tangent construction much like that we find here is also found in phase separation in {\it disordered} active systems (e.g., ``Motility Induced Phase Separation" (``MIPS") \cite{mipsreview}). It thus appears that this uncommon construction is, paradoxically, quite common in active systems.

\vspace{.2in}

\section{Conclusion}

We have shown that a dry active autochemotaxic flock, or, more generally, {\it any} flock with { sufficiently strong} attractive interactions between the flockers, can become unstable to the formation of density bands parallel to the mean direction of motion of the flock. { The attractive interactions must drive the inverse compressibiliy negative.} This behavior may be connected to the formation of, e.g., unidirectional ant trails. The instability is caused by the autochemotaxic ``pressure" or other attractive mechanism overcoming the usual repulsive isotropic pressure, causing the inverse compressibility $B$ of the flock to be negative. We investigated that instability, and specifically demonstrated that:

\begin{itemize}
    \item The instability is anisotropic.  The direction of maximum growth rate is \textit{perpendicular} to the mean flock motion. This initially causes the formation of a growing wavelike modulation of the local density \textit{parallel} to mean flock motion.  The maximum growth rate ${\rm Im}~\omega_{max} \propto \sqrt{|B|}$.
    
    \item This modulation quickly grows into a ``phase separated'' flock consisting of high density plateaus separated by low density plateaus. The boundaries of the plateaus are weakly attracted to one another. Our analytical results predict the final state of the system will consist of a single high density plateau and a single low density plateau.
    
    \item  The approach to this final state is extremely (exponentially) slow in our 1D numerical solutions. However, this  one dimensional picture is almost certainly incorrect for the late stage dynamics. We believe that noise, and the wandering of the band boundaries noise induces, will considerably accelerate  these coarsening dynamics. In particular, two high density bands may collide and ``zip up'', as illustrated in figure \ref{noiseBands}. We speculate that this process will make the time required to reach the final state of one high density band ``coexisting" with one low density band grow algebraically, rather than exponentially, with system size. 
    
    \item Sufficiently near the ``critical point" of this non-equilibrium phase separation, the coexisting plateau densities can be determined by a ``common tangent construction" quite like that used in equilibrium phase separation. Further from the critical point, however, this becomes an ``{\it un}common tangent construction", with the effective free energy having the same slope at both plateau densities, but with the tangents at those two densities being distinct, as illustrated in figure \ref{UTCDemo}. 
    
    \item Our system exhibits both binodal and spinodal lines, just as in equilibrium phase separation, as illustrated in figure \ref{spinodal}.
    
    \item We have analytically determined the plateau densities, and the profile of both the density and the velocity at the interfaces between plateaus, in a perturbation theory asymptotically valid close to the critical point. This analytic theory compares extremely well with our long-time numerical solutions, even far from the critical point.
    
\end{itemize}

Our work has focused on a linear analysis of a noiseless,  polar ordered, dry, active autochemotaxic flock. We put forward two topics of future work. The first is on the inclusion of fluctuation effects,  both in the dynamics of the approach to the final phase separated state, and in the determination of that state as well. The second is developing a theory of flock interface dynamics to act as the cornerstone for understanding the late stage phase separation dynamics.

\section{Acknowledgements}
We thank Tristan Ursell and Justin Kittel for stimulating discussions of their experiments on the dynamics of ant trails; these discussions motivated our work. We'd also like to thank the Max Planck Institute for the Physics of Complex Systems (MPIKPS), Dresden,  for their hospitality during a portion of this work. We would also like to acknowledge Ants for their interesting behaviors. Maxx Miller thanks Isabelle and Totoro Kloc  for their support while this work was underway.

\bibliographystyle{apsrev4-1}

\begin{thebibliography}{35}%
\makeatletter
\providecommand \@ifxundefined [1]{%
 \@ifx{#1\undefined}
}%
\providecommand \@ifnum [1]{%
 \ifnum #1\expandafter \@firstoftwo
 \else \expandafter \@secondoftwo
 \fi
}%
\providecommand \@ifx [1]{%
 \ifx #1\expandafter \@firstoftwo
 \else \expandafter \@secondoftwo
 \fi
}%
\providecommand \natexlab [1]{#1}%
\providecommand \enquote  [1]{``#1''}%
\providecommand \bibnamefont  [1]{#1}%
\providecommand \bibfnamefont [1]{#1}%
\providecommand \citenamefont [1]{#1}%
\providecommand \href@noop [0]{\@secondoftwo}%
\providecommand \href [0]{\begingroup \@sanitize@url \@href}%
\providecommand \@href[1]{\@@startlink{#1}\@@href}%
\providecommand \@@href[1]{\endgroup#1\@@endlink}%
\providecommand \@sanitize@url [0]{\catcode `\\12\catcode `\$12\catcode
  `\&12\catcode `\#12\catcode `\^12\catcode `\_12\catcode `\%12\relax}%
\providecommand \@@startlink[1]{}%
\providecommand \@@endlink[0]{}%
\providecommand \url  [0]{\begingroup\@sanitize@url \@url }%
\providecommand \@url [1]{\endgroup\@href {#1}{\urlprefix }}%
\providecommand \urlprefix  [0]{URL }%
\providecommand \Eprint [0]{\href }%
\providecommand \doibase [0]{http://dx.doi.org/}%
\providecommand \selectlanguage [0]{\@gobble}%
\providecommand \bibinfo  [0]{\@secondoftwo}%
\providecommand \bibfield  [0]{\@secondoftwo}%
\providecommand \translation [1]{[#1]}%
\providecommand \BibitemOpen [0]{}%
\providecommand \bibitemStop [0]{}%
\providecommand \bibitemNoStop [0]{.\EOS\space}%
\providecommand \EOS [0]{\spacefactor3000\relax}%
\providecommand \BibitemShut [1]{\csname bibitem#1\endcsname}%
\let\auto@bib@innerbib\@empty
\bibitem [{\citenamefont {Vicsek}\ \emph {et~al.}(1995)\citenamefont {Vicsek},
  \citenamefont {Czir\'ok}, \citenamefont {Ben-Jacob}, \citenamefont {Cohen},\
  and\ \citenamefont {Shochet}}]{Vicsek}%
  \BibitemOpen
  \bibfield  {author} {\bibinfo {author} {\bibfnamefont {T.}~\bibnamefont
  {Vicsek}}, \bibinfo {author} {\bibfnamefont {A.}~\bibnamefont {Czir\'ok}},
  \bibinfo {author} {\bibfnamefont {E.}~\bibnamefont {Ben-Jacob}}, \bibinfo
  {author} {\bibfnamefont {I.}~\bibnamefont {Cohen}}, \ and\ \bibinfo {author}
  {\bibfnamefont {O.}~\bibnamefont {Shochet}},\ }\href {\doibase
  10.1103/PhysRevLett.75.1226} {\bibfield  {journal} {\bibinfo  {journal}
  {Phys. Rev. Lett.}\ }\textbf {\bibinfo {volume} {75}},\ \bibinfo {pages}
  {1226} (\bibinfo {year} {1995})}\BibitemShut{NoStop}%
\bibitem [{\citenamefont {Toner}\ and\ \citenamefont {Tu}(1995)}]{TT1}%
  \BibitemOpen
  \bibfield  {author} {\bibinfo {author} {\bibfnamefont {J.}~\bibnamefont
  {Toner}}\ and\ \bibinfo {author} {\bibfnamefont {Y.}~\bibnamefont {Tu}},\
  }\href {\doibase 10.1103/PhysRevLett.75.4326} {\bibfield  {journal} {\bibinfo
   {journal} {Phys. Rev. Lett.}\ }\textbf {\bibinfo {volume} {75}},\ \bibinfo
  {pages} {4326} (\bibinfo {year} {1995})}\BibitemShut{NoStop}%
\bibitem [{\citenamefont {Tu}\ \emph {et~al.}(1998)\citenamefont {Tu},
  \citenamefont {Toner},\ and\ \citenamefont {Ulm}}]{TT2}%
  \BibitemOpen
  \bibfield  {author} {\bibinfo {author} {\bibfnamefont {Y.}~\bibnamefont
  {Tu}}, \bibinfo {author} {\bibfnamefont {J.}~\bibnamefont {Toner}}, \ and\
  \bibinfo {author} {\bibfnamefont {M.}~\bibnamefont {Ulm}},\ }\href {\doibase
  10.1103/PhysRevLett.80.4819} {\bibfield  {journal} {\bibinfo  {journal}
  {Phys. Rev. Lett.}\ }\textbf {\bibinfo {volume} {80}},\ \bibinfo {pages}
  {4819} (\bibinfo {year} {1998})}\BibitemShut{NoStop}%
\bibitem [{\citenamefont {Toner}\ and\ \citenamefont {Tu}(1998)}]{TT3}%
  \BibitemOpen
  \bibfield  {author} {\bibinfo {author} {\bibfnamefont {J.}~\bibnamefont
  {Toner}}\ and\ \bibinfo {author} {\bibfnamefont {Y.}~\bibnamefont {Tu}},\
  }\href {\doibase 10.1103/PhysRevE.58.4828} {\bibfield  {journal} {\bibinfo
  {journal} {Phys. Rev. E}\ }\textbf {\bibinfo {volume} {58}},\ \bibinfo
  {pages} {4828} (\bibinfo {year} {1998})}\BibitemShut{NoStop}%
\bibitem [{\citenamefont {Toner}\ \emph {et~al.}(2005)\citenamefont {Toner},
  \citenamefont {Tu},\ and\ \citenamefont {Ramaswamy}}]{TT4}%
  \BibitemOpen
  \bibfield  {author} {\bibinfo {author} {\bibfnamefont {J.}~\bibnamefont
  {Toner}}, \bibinfo {author} {\bibfnamefont {Y.}~\bibnamefont {Tu}}, \ and\
  \bibinfo {author} {\bibfnamefont {S.}~\bibnamefont {Ramaswamy}},\ }\href
  {\doibase https://doi.org/10.1016/j.aop.2005.04.011} {\bibfield  {journal}
  {\bibinfo  {journal} {Annals of Physics}\ }\textbf {\bibinfo {volume}
  {318}},\ \bibinfo {pages} {170} (\bibinfo {year} {2005})},\ \bibinfo {note}
  {\uppercase{S}pecial Issue}\BibitemShut{NoStop}%
\bibitem [{\citenamefont {Pline}\ \emph {et~al.}(1988)\citenamefont {Pline},
  \citenamefont {Diez},\ and\ \citenamefont {Dusenbery}}]{Thermotaxis}%
  \BibitemOpen
  \bibfield  {author} {\bibinfo {author} {\bibfnamefont {M.}~\bibnamefont
  {Pline}}, \bibinfo {author} {\bibfnamefont {J.~A.}\ \bibnamefont {Diez}}, \
  and\ \bibinfo {author} {\bibfnamefont {D.~B.}\ \bibnamefont {Dusenbery}},\
  }\href@noop {} {\bibfield  {journal} {\bibinfo  {journal} {J. Nematology}\
  }\textbf {\bibinfo {volume} {20}},\ \bibinfo {pages} {605} (\bibinfo {year}
  {1988})}\BibitemShut{NoStop}%
\bibitem [{\citenamefont {Anderson}\ and\ \citenamefont
  {Mackie}(1977)}]{Phototaxis}%
  \BibitemOpen
  \bibfield  {author} {\bibinfo {author} {\bibfnamefont {P.~A.}\ \bibnamefont
  {Anderson}}\ and\ \bibinfo {author} {\bibfnamefont {G.~O.}\ \bibnamefont
  {Mackie}},\ }\href@noop {} {\bibfield  {journal} {\bibinfo  {journal}
  {Science}\ }\textbf {\bibinfo {volume} {197}},\ \bibinfo {pages} {186}
  (\bibinfo {year} {1977})}\BibitemShut{NoStop}%
\bibitem [{\citenamefont {Stock}\ and\ \citenamefont
  {Baker}()}]{chemotaxisExample}%
  \BibitemOpen
  \bibfield  {author} {\bibinfo {author} {\bibfnamefont {J.}~\bibnamefont
  {Stock}}\ and\ \bibinfo {author} {\bibfnamefont {M.}~\bibnamefont {Baker}},\
  }in\ \href {\doibase https://doi.org/10.1016/B978-012373944-5.00068-7} {\emph
  {\bibinfo {booktitle} {Encyclopedia of Microbiology (Third Edition)}}},\
  \bibinfo {editor} {edited by\ \bibinfo {editor} {\bibfnamefont
  {M.}~\bibnamefont {Schaechter}}},\ \bibinfo {note} {(Academic Press, Oxford,
  2009), pp.71-78}\BibitemShut{NoStop}%
\bibitem [{\citenamefont {Mesibov}\ and\ \citenamefont {Adler}(1972)}]{Ecoli}%
  \BibitemOpen
  \bibfield  {author} {\bibinfo {author} {\bibfnamefont {R.}~\bibnamefont
  {Mesibov}}\ and\ \bibinfo {author} {\bibfnamefont {J.}~\bibnamefont
  {Adler}},\ }\href@noop {} {\bibfield  {journal} {\bibinfo  {journal} {J.
  Bacteriology}\ }\textbf {\bibinfo {volume} {112}},\ \bibinfo {pages} {315}
  (\bibinfo {year} {1972})}\BibitemShut{NoStop}%
\bibitem [{\citenamefont {Doane}\ and\ \citenamefont {McManus}()}]{Moth}%
  \BibitemOpen
  \bibinfo {editor} {\bibfnamefont {C.~C.}\ \bibnamefont {Doane}}\ and\
  \bibinfo {editor} {\bibfnamefont {M.~L.}\ \bibnamefont {McManus}},\ eds.,\
  \href@noop {} {\emph {\bibinfo {title} {The Gypsy moth : research toward
  integrated pest management}}},\ \bibinfo {note} {(U.S. De- partment of
  Agriculture, Washington, D.C., 1981).}\BibitemShut{Stop}%
\bibitem [{\citenamefont {Hildebrand}\ and\ \citenamefont
  {Kaupp}(2005)}]{Taxis1}%
  \BibitemOpen
  \bibfield  {author} {\bibinfo {author} {\bibfnamefont {E.}~\bibnamefont
  {Hildebrand}}\ and\ \bibinfo {author} {\bibfnamefont {U.~B.}\ \bibnamefont
  {Kaupp}},\ }\href@noop {} {\bibfield  {journal} {\bibinfo  {journal} {Ann. N.
  Y. Acad. Sci.}\ }\textbf {\bibinfo {volume} {1061}},\ \bibinfo {pages} {221}
  (\bibinfo {year} {2005})}\BibitemShut{NoStop}%
\bibitem [{\citenamefont {Topaz}\ \emph {et~al.}(2012)\citenamefont {Topaz},
  \citenamefont {D'Orsogna}, \citenamefont {Edelstein-Keshet},\ and\
  \citenamefont {Bernoff}}]{Locust}%
  \BibitemOpen
  \bibfield  {author} {\bibinfo {author} {\bibfnamefont {C.~M.}\ \bibnamefont
  {Topaz}}, \bibinfo {author} {\bibfnamefont {M.~R.}\ \bibnamefont
  {D'Orsogna}}, \bibinfo {author} {\bibfnamefont {L.}~\bibnamefont
  {Edelstein-Keshet}}, \ and\ \bibinfo {author} {\bibfnamefont {A.~J.}\
  \bibnamefont {Bernoff}},\ }\href@noop {} {\bibfield  {journal} {\bibinfo
  {journal} {PLoS Comput. Biol.}\ }\textbf {\bibinfo {volume} {8}},\ \bibinfo
  {pages} {e1002642} (\bibinfo {year} {2012})}\BibitemShut{NoStop}%
\bibitem [{\citenamefont {Couzin}\ and\ \citenamefont {Franks}(2003)}]{Ants1}%
  \BibitemOpen
  \bibfield  {author} {\bibinfo {author} {\bibfnamefont {I.~D.}\ \bibnamefont
  {Couzin}}\ and\ \bibinfo {author} {\bibfnamefont {N.~R.}\ \bibnamefont
  {Franks}},\ }\href@noop {} {\bibfield  {journal} {\bibinfo  {journal} {Proc.
  Biol. Sci.}\ }\textbf {\bibinfo {volume} {270}},\ \bibinfo {pages} {139}
  (\bibinfo {year} {2003})}\BibitemShut{NoStop}%
\bibitem [{\citenamefont {Dussutour}\ \emph {et~al.}(2009)\citenamefont
  {Dussutour}, \citenamefont {Beekman}, \citenamefont {Nicolis},\ and\
  \citenamefont {Meyer}}]{AntNoise1}%
  \BibitemOpen
  \bibfield  {author} {\bibinfo {author} {\bibfnamefont {A.}~\bibnamefont
  {Dussutour}}, \bibinfo {author} {\bibfnamefont {M.}~\bibnamefont {Beekman}},
  \bibinfo {author} {\bibfnamefont {S.~C.}\ \bibnamefont {Nicolis}}, \ and\
  \bibinfo {author} {\bibfnamefont {B.}~\bibnamefont {Meyer}},\ }\href@noop {}
  {\bibfield  {journal} {\bibinfo  {journal} {Proc. Biol. Sci.}\ }\textbf
  {\bibinfo {volume} {276}},\ \bibinfo {pages} {4353} (\bibinfo {year}
  {2009})}\BibitemShut{NoStop}%
\bibitem [{\citenamefont {Keller}\ and\ \citenamefont
  {Segel}(1971{\natexlab{a}})}]{KS1}%
  \BibitemOpen
  \bibfield  {author} {\bibinfo {author} {\bibfnamefont {E.~F.}\ \bibnamefont
  {Keller}}\ and\ \bibinfo {author} {\bibfnamefont {L.~A.}\ \bibnamefont
  {Segel}},\ }\href {\doibase https://doi.org/10.1016/0022-5193(71)90050-6}
  {\bibfield  {journal} {\bibinfo  {journal} {Journal of Theoretical Biology}\
  }\textbf {\bibinfo {volume} {30}},\ \bibinfo {pages} {225} (\bibinfo {year}
  {1971}{\natexlab{a}})}\BibitemShut{NoStop}%
\bibitem [{\citenamefont {Keller}\ and\ \citenamefont {Segel}(1970)}]{KS2}%
  \BibitemOpen
  \bibfield  {author} {\bibinfo {author} {\bibfnamefont {E.~F.}\ \bibnamefont
  {Keller}}\ and\ \bibinfo {author} {\bibfnamefont {L.~A.}\ \bibnamefont
  {Segel}},\ }\href {\doibase https://doi.org/10.1016/0022-5193(70)90092-5}
  {\bibfield  {journal} {\bibinfo  {journal} {Journal of Theoretical Biology}\
  }\textbf {\bibinfo {volume} {26}},\ \bibinfo {pages} {399} (\bibinfo {year}
  {1970})}\BibitemShut{NoStop}%
\bibitem [{\citenamefont {Keller}\ and\ \citenamefont
  {Segel}(1971{\natexlab{b}})}]{KS3}%
  \BibitemOpen
  \bibfield  {author} {\bibinfo {author} {\bibfnamefont {E.~F.}\ \bibnamefont
  {Keller}}\ and\ \bibinfo {author} {\bibfnamefont {L.~A.}\ \bibnamefont
  {Segel}},\ }\href {\doibase https://doi.org/10.1016/0022-5193(71)90051-8}
  {\bibfield  {journal} {\bibinfo  {journal} {Journal of Theoretical Biology}\
  }\textbf {\bibinfo {volume} {30}},\ \bibinfo {pages} {235} (\bibinfo {year}
  {1971}{\natexlab{b}})}\BibitemShut{NoStop}%
\bibitem [{\citenamefont {Horstmann}(2003)}]{KSReview}%
  \BibitemOpen
  \bibfield  {author} {\bibinfo {author} {\bibfnamefont {D.}~\bibnamefont
  {Horstmann}},\ }\href@noop {} {\bibfield  {journal} {\bibinfo  {journal}
  {Jahresbericht der Deutschen Mathematiker-Vereinigung}\ }\textbf {\bibinfo
  {volume} {105}},\ \bibinfo {pages} {103} (\bibinfo {year}
  {2003})}\BibitemShut{NoStop}%
\bibitem [{\citenamefont {Stark}(2018)}]{KSCollective}%
  \BibitemOpen
  \bibfield  {author} {\bibinfo {author} {\bibfnamefont {H.}~\bibnamefont
  {Stark}},\ }\href@noop {} {\bibfield  {journal} {\bibinfo  {journal} {Acc.
  Chem. Res.}\ }\textbf {\bibinfo {volume} {51}},\ \bibinfo {pages} {2681}
  (\bibinfo {year} {2018})}\BibitemShut{NoStop}%
\bibitem [{\citenamefont {Mermin}\ and\ \citenamefont {Wagner}(1966)}]{MW}%
  \BibitemOpen
  \bibfield  {author} {\bibinfo {author} {\bibfnamefont {N.~D.}\ \bibnamefont
  {Mermin}}\ and\ \bibinfo {author} {\bibfnamefont {H.}~\bibnamefont
  {Wagner}},\ }\href {\doibase 10.1103/PhysRevLett.17.1133} {\bibfield
  {journal} {\bibinfo  {journal} {Phys. Rev. Lett.}\ }\textbf {\bibinfo
  {volume} {17}},\ \bibinfo {pages} {1133} (\bibinfo {year}
  {1966})}\BibitemShut{NoStop}%
\bibitem [{\citenamefont {Dormann}\ and\ \citenamefont
  {Weijer}(2006)}]{Taxis3}%
  \BibitemOpen
  \bibfield  {author} {\bibinfo {author} {\bibfnamefont {D.}~\bibnamefont
  {Dormann}}\ and\ \bibinfo {author} {\bibfnamefont {C.~J.}\ \bibnamefont
  {Weijer}},\ }\href {\doibase https://doi.org/10.1016/j.gde.2006.06.003}
  {\bibfield  {journal} {\bibinfo  {journal} {Current Opinion in Genetics \&
  Development}\ }\textbf {\bibinfo {volume} {16}},\ \bibinfo {pages} {367}
  (\bibinfo {year} {2006})}\BibitemShut{NoStop}%
\bibitem [{\citenamefont {Toner}(2012{\natexlab{a}})}]{rean}%
  \BibitemOpen
  \bibfield  {author} {\bibinfo {author} {\bibfnamefont {J.}~\bibnamefont
  {Toner}},\ }\href {\doibase 10.1103/PhysRevE.86.031918} {\bibfield  {journal}
  {\bibinfo  {journal} {Phys. Rev. E}\ }\textbf {\bibinfo {volume} {86}},\
  \bibinfo {pages} {031918} (\bibinfo {year} {2012}{\natexlab{a}})}\BibitemShut{NoStop}%
\bibitem [{\citenamefont {Bertrand}\ and\ \citenamefont {Lee}(2022)}]{BLee}%
  \BibitemOpen
  \bibfield  {author} {\bibinfo {author} {\bibfnamefont {T.}~\bibnamefont
  {Bertrand}}\ and\ \bibinfo {author} {\bibfnamefont {C.~F.}\ \bibnamefont
  {Lee}},\ }\href {\doibase 10.1103/PhysRevResearch.4.L022046} {\bibfield
  {journal} {\bibinfo  {journal} {Phys. Rev. Res.}\ }\textbf {\bibinfo {volume}
  {4}},\ \bibinfo {pages} {L022046} (\bibinfo {year} {2022})}\BibitemShut{NoStop}%
\bibitem [{\citenamefont {Bertin}\ \emph {et~al.}(2009)\citenamefont {Bertin},
  \citenamefont {Droz},\ and\ \citenamefont {Grégoire}}]{banding1}%
  \BibitemOpen
  \bibfield  {author} {\bibinfo {author} {\bibfnamefont {E.}~\bibnamefont
  {Bertin}}, \bibinfo {author} {\bibfnamefont {M.}~\bibnamefont {Droz}}, \ and\
  \bibinfo {author} {\bibfnamefont {G.}~\bibnamefont {Grégoire}},\ }\href
  {\doibase 10.1088/1751-8113/42/44/445001} {\bibfield  {journal} {\bibinfo
  {journal} {Journal of Physics A: Mathematical and Theoretical}\ }\textbf
  {\bibinfo {volume} {42}},\ \bibinfo {pages} {445001} (\bibinfo {year}
  {2009})}, \BibitemShut{NoStop}%
\bibitem [{\citenamefont {Mishra}\ \emph {et~al.}(2010)\citenamefont {Mishra},
  \citenamefont {Baskaran},\ and\ \citenamefont {Marchetti}}]{Banding2}%
  \BibitemOpen
  \bibfield  {author} {\bibinfo {author} {\bibfnamefont {S.}~\bibnamefont
  {Mishra}}, \bibinfo {author} {\bibfnamefont {A.}~\bibnamefont {Baskaran}}, \
  and\ \bibinfo {author} {\bibfnamefont {M.~C.}\ \bibnamefont {Marchetti}},\
  }\href {\doibase 10.1103/PhysRevE.81.061916} {\bibfield  {journal} {\bibinfo
  {journal} {Phys. Rev. E}\ }\textbf {\bibinfo {volume} {81}},\ \bibinfo
  {pages} {061916} (\bibinfo {year} {2010})}\BibitemShut{NoStop}%
\bibitem [{\citenamefont {Kardar}\ \emph {et~al.}(1986)\citenamefont {Kardar},
  \citenamefont {Parisi},\ and\ \citenamefont {Zhang}}]{KPZ1}%
  \BibitemOpen
  \bibfield  {author} {\bibinfo {author} {\bibfnamefont {M.}~\bibnamefont
  {Kardar}}, \bibinfo {author} {\bibfnamefont {G.}~\bibnamefont {Parisi}}, \
  and\ \bibinfo {author} {\bibfnamefont {Y.-C.}\ \bibnamefont {Zhang}},\ }\href
  {\doibase 10.1103/PhysRevLett.56.889} {\bibfield  {journal} {\bibinfo
  {journal} {Phys. Rev. Lett.}\ }\textbf {\bibinfo {volume} {56}},\ \bibinfo
  {pages} {889} (\bibinfo {year} {1986})}\BibitemShut{NoStop}%
\bibitem [{\citenamefont {Chaikin}\ and\ \citenamefont
  {Lubensky}(1995)}]{chaikin}%
  \BibitemOpen
  \bibfield  {author} {\bibinfo {author} {\bibfnamefont {P.~M.}\ \bibnamefont
  {Chaikin}}\ and\ \bibinfo {author} {\bibfnamefont {T.~C.}\ \bibnamefont
  {Lubensky}},\ }\href {\doibase 10.1017/CBO9780511813467} {\emph {\bibinfo
  {title} {Principles of Condensed Matter Physics}}}\ (\bibinfo  {publisher}
  {Cambridge University Press},\ \bibinfo {year} {1995})\BibitemShut{NoStop}%
\bibitem [{\citenamefont {Cates}\ and\ \citenamefont
  {Tailleur}(2015)}]{mipsreview}%
  \BibitemOpen
  \bibfield  {author} {\bibinfo {author} {\bibfnamefont {M.~E.}\ \bibnamefont
  {Cates}}\ and\ \bibinfo {author} {\bibfnamefont {J.}~\bibnamefont
  {Tailleur}},\ }\href {\doibase 10.1146/annurev-conmatphys-031214-014710}
  {\bibfield  {journal} {\bibinfo  {journal} {Annual Review of Condensed Matter
  Physics}\ }\textbf {\bibinfo {volume} {6}},\ \bibinfo {pages} {219} (\bibinfo
  {year} {2015})}\BibitemShut{NoStop}%
\bibitem [{\citenamefont {Ma}(1995)}]{Ma}%
  \BibitemOpen
  \bibfield  {author} {\bibinfo {author} {\bibfnamefont {S.~K.}\ \bibnamefont
  {Ma}},\ }\href@noop {} {\emph {\bibinfo {title} {Modern Theory of Critical
  Phenomena}}}\ (\bibinfo {address} {Benjamin, Reading, Mass},\ \bibinfo {year}
  {1995})\BibitemShut{NoStop}%
\bibitem [{\citenamefont {Toner}(2012{\natexlab{b}})}]{TT5}%
  \BibitemOpen
  \bibfield  {author} {\bibinfo {author} {\bibfnamefont {J.}~\bibnamefont
  {Toner}},\ }\href {\doibase 10.1103/PhysRevLett.108.088102} {\bibfield
  {journal} {\bibinfo  {journal} {Phys. Rev. Lett.}\ }\textbf {\bibinfo
  {volume} {108}},\ \bibinfo {pages} {088102} (\bibinfo {year}
  {2012}{\natexlab{b}})}\BibitemShut{NoStop}%
\bibitem [{\citenamefont {Chen}\ \emph {et~al.}(2020)\citenamefont {Chen},
  \citenamefont {Lee},\ and\ \citenamefont {Toner}}]{Malthus}%
  \BibitemOpen
  \bibfield  {author} {\bibinfo {author} {\bibfnamefont {L.}~\bibnamefont
  {Chen}}, \bibinfo {author} {\bibfnamefont {C.~F.}\ \bibnamefont {Lee}}, \
  and\ \bibinfo {author} {\bibfnamefont {J.}~\bibnamefont {Toner}},\ }\href
  {\doibase 10.1103/PhysRevLett.125.098003} {\bibfield  {journal} {\bibinfo
  {journal} {Phys. Rev. Lett.}\ }\textbf {\bibinfo {volume} {125}},\ \bibinfo
  {pages} {098003} (\bibinfo {year} {2020})}\BibitemShut{NoStop}%
\bibitem [{\citenamefont {Hillen}\ and\ \citenamefont
  {Painter}(2009)}]{HillenPainterCompsci}%
  \BibitemOpen
  \bibfield  {author} {\bibinfo {author} {\bibfnamefont {T.}~\bibnamefont
  {Hillen}}\ and\ \bibinfo {author} {\bibfnamefont {K.~J.}\ \bibnamefont
  {Painter}},\ }\href@noop {} {\bibfield  {journal} {\bibinfo  {journal}
  {Journal of Mathematical Biology}\ }\textbf {\bibinfo {volume} {58}},\
  \bibinfo {pages} {183} (\bibinfo {year} {2009})}\BibitemShut{NoStop}%
\bibitem [{Tru()}]{TrueVelocity}%
  \BibitemOpen
  \href@noop {} {}\bibinfo {note} { The velocity field in our field equations is
  \textit{not} the ``true'', or ``full'', velocity field. The true velocity
  field is defined as the velocity that satisfies $\mathbf{J}=\rho \mathbf{v}$.
  The velocity field that we employ is, instead, a proxy for the polarization
  field of the flock $\mathbf{ \hp}$, defined as $\mathbf{v}=v_0\hat{p}$. Where
  $\mathbf{ \hp}$ is the local ``polarization''; that is, a unit vector along
  the direction in which the local flockers are pointing, and $v_0$ is the
  speed of the boids in a uniform state. By ``pointing'' here, we mean the
  direction along which the ``motors'' with which the flockers are propelling
  themselves are acting. Because there are many other effects that can lead to
  a net current of the flockers aside from their self-propulsion (e.g.,
  pressure forces, Brownian noise, etc.), the net current $\bj$ can get
  additional contributions that are \textit{ not} simply given by $\rho\bv$,
  \textit{with this definition of} $\bv$. This fact then requires that we
  include the symmetry allowed $k_1$, $k_{1a}$, $k_2$ and $k_{2a}$ terms in the
  equation of motion III.2. Note that even if we \textit{had} defined the
  velocity field via $\mathbf{J}=\rho \mathbf{v}$, that exact relation would
  still have broken down once we coarse grained, since coarse graining means
  that we have averaged out some of the short-wavelength components of the
  velocity. The remaining components therefore no longer constitute the full
  velocity field, and so, in general, will not obey $\mathbf{J}=\rho
  \mathbf{v}$. Operationally, this fact is manifested by the generation of the
  $k_1$, $k_{1a}$, $k_2$ and $k_{2a}$ terms by coarse graining (i.e., by the
  dynamical RG), even if those terms are absent in the original (bare)
  model}\BibitemShut{NoStop}%
\bibitem [{\citenamefont {Miller}\ and\ \citenamefont {Toner}()}]{ASP}%
  \BibitemOpen
  \bibfield  {author} {\bibinfo {author} {\bibfnamefont {M.}~\bibnamefont
  {Miller}}\ and\ \bibinfo {author} {\bibfnamefont {J.}~\bibnamefont {Toner}},\
  }\href@noop {} {\ }\bibinfo {note} {The associated short paper}\BibitemShut{NoStop}%
\bibitem [{w2f()}]{w2foot}%
  \BibitemOpen
  \href@noop {} {}\bibinfo {note} {Strictly speaking, we should also have
  included a term proportional to $\pp_x\Delta\rho^3$on the right hand side of
  the equation of motion for $\rho$. Since we are looking for solutions that
  are independent of $x$, this term vanishes anyway.}\BibitemShut{Stop}%
\end{thebibliography}
\end{document}